\documentclass{desyprocA4}
\usepackage{color}
\usepackage{parskip}               
\usepackage{epsfig}
\usepackage{amssymb}
\usepackage{subfigure}
\usepackage[ , ,bf,it]{caption}    
\newcommand\pubdate{\today}
\def\eslt{E_T^{\rm miss}}

\def\esl{\not\!\!{E}}

\def\to{\rightarrow}

\def\bi{\begin{itemize}}
 \def\ei{\end{itemize}}
\def\te{\tilde e}

\def\c1p{C1^\prime}
\def\msq3{\overline{m}_{\tilde{q}}(3)}
\def\ta{\tilde a}
\def\tG{\tilde G}

\def\tl{\tilde l}

\def\ta{\tilde a}

\def\tb{\tilde b}

\def\tst{\tilde t}
\def\ttau{\tilde \tau}
\def\tmu{\tilde \mu}
\def\tg{\tilde g}
\def\tnu{\tilde\nu}
\def\tell{\tilde\ell}
\def\tq{\tilde q}

\def\be{\begin{equation}}  
\def\ee{\end{equation}}  
\def\bea{\begin{eqnarray}}  
\def\eea{\end{eqnarray}}  
\def\tw{\tilde\chi}
\def\twp{\tilde\chi^+}
\def\twm{\tilde\chi^-}
\def\twpm{\tilde\chi^\pm}
\def\tz{\tilde\chi^0}
\newcommand{\alt}{\mbox{$\;\raisebox{-1mm}{$\stackrel{\scriptstyle<}{\scriptstyle\sim}$}\;$}}
\newcommand{\agt}{\mbox{$\;\raisebox{-1mm}{$\stackrel{\scriptstyle>}{\scriptstyle\sim}$}\;$}}

\def\beq{\begin{equation}}
\def\eeq#1{\label{#1}\end{equation}}
\def\eeqn{\end{equation}}

\newenvironment{Eqnarray}%
   {\arraycolsep 0.14em\begin{eqnarray}}{\end{eqnarray}}
\def\beqa{\begin{Eqnarray}}
\def\eeqa#1{\label{#1}\end{Eqnarray}}
\def\eeqan{\end{Eqnarray}}

\newcommand{\ra}            {\ensuremath{ \rightarrow     }}

\begin{document}
\title{Post LHC8 SUSY benchmark points for ILC physics}

\author{{\slshape Howard Baer$^1$, Jenny List$^2$}\\[1ex]
$^1$University of Oklahoma, Norman, OK 73019, USA\\
$^2$DESY, Notkestra{\ss}e 85, 22607 Hamburg, Germany\\
}

\desyproc{DESY 13-120}

\maketitle

\pubdate

\begin{abstract}
We re-evaluate prospects for supersymmetry at the proposed 
International Linear $e^+e^-$ Collider (ILC) in light of the first
two years of serious data taking at LHC: LHC7 with $\sim 5\,$fb$^{-1}$ of $pp$ collisions at $\sqrt{s}=7\,$TeV 
and LHC8 with $\sim 20\,$fb$^{-1}$ at $\sqrt{s}=8\,$TeV.
Strong new limits from LHC8 SUSY searches, along with the discovery of a Higgs boson
with $m_h\simeq 125\,$GeV, suggest a paradigm shift from previously popular models
to ones with new and compelling signatures. 
After a review of the current status of supersymmetry, 
we present a variety of new ILC benchmark models, including: 
natural SUSY, 
radiatively-driven natural SUSY (RNS), 
NUHM2 with low $m_A$, 
a focus point case from mSUGRA/CMSSM, 
non-universal gaugino mass (NUGM) model, 
$\ttau$-coannihilation, 
Kallosh-Linde/spread SUSY model, 
mixed gauge-gravity mediation,
normal scalar mass hierarchy (NMH), 
and one example with the recently discovered Higgs boson being the heavy $CP$-even state $H$.
While all these models at present elude the latest LHC8 limits, 
they do offer intriguing case study possibilities for ILC operating at $\sqrt{s}\simeq 0.25-1\,$TeV.
The benchmark points also present a view of the widely diverse SUSY phenomena 
which might still be expected in the post LHC8 era at both LHC and ILC.
\end{abstract}

\section{Introduction}

\subsection{Motivation}

Supersymmetry (SUSY) is a quantum spacetime symmetry which predicts a correspondence between
bosonic and fermionic fields~\cite{Wess:1973kz,Salam:1974jj,Salam:1974ig,Baer:2006rs}. 
Supersymmetry is particularly appealing for theories of
particle physics in that it reduces scalar field quadratic divergences to merely logarithmic. 
This fact allows for an elegant solution to the notorious gauge hierarchy problem, rendering
the weak scale stable against quantum corrections and allowing for stable extrapolations of the
Standard Model (SM) into the far ultraviolet ($E\gg M_{\mathrm{weak}}$) regime~\cite{Witten:1981nf,Kaul:1981wp}.
Thus, SUSY provides an 
avenue for connecting the Standard Model to ideas of grand unification (GUTs) and/or string theory,
and provides a route to unification with gravity via local SUSY, or supergravity 
theories~\cite{Ferrara:1976kg,Cremmer:1982en,Nilles:1983ge}. 

While models of weak scale supersymmetry are theoretically compelling, we note here 
that a variety of indirect evidence from experiment has emerged which provides
support for the idea of weak scale SUSY:
\begin{itemize}
\item {\it Gauge coupling unification:} The values of the three SM gauge couplings,
measured at energy scale $Q\simeq M_Z$ at the CERN LEP collider, 
when extrapolated to high energy scales via renormalization group (RG) running in the
Minimal Supersymmetric Standard Model (MSSM)~\cite{Dimopoulos:1981yj}, very nearly meet at a point 
around $Q\simeq 2\times 10^{16}\,$GeV~\cite{Amaldi:1991cn,Langacker:1991an,Ellis:1990wk}. 
Unification of gauge couplings is predicted by GUTs and
string theories. Gauge coupling unification is violated by numerous standard deviations 
under SM RG running.
\item {\it Precision electroweak measurements:} Fits of precision electroweak
observables (EWPO)  to SUSY model predictions find accord provided there exists a rather heavy SUSY 
particle mass spectrum~\cite{Heinemeyer:2006px}. 
Meanwhile, models such as minimal technicolor are highly stressed if not ruled
out by EWPO.
\item {\it Top quark mass and electroweak symmetry breaking:} The electroweak scalar potential is highly
constrained in SUSY theories compared to the SM, and it is not immediately clear if electroweak symmetry
can be properly broken, yielding the required vector boson and fermion masses while leaving the 
photon massless. 
In top-down theories, the soft breaking Higgs mass $m_{H_u}^2$ is driven to negative values
by the large top quark Yukawa coupling, triggering an appropriate breakdown of EW symmetry, 
provided that the top quark mass $m_t\simeq 150-200\,$GeV~\cite{Ibanez:1982fr,Inoue:1982pi,Ellis:1983bp,AlvarezGaume:1983gj}. 
The latest measurements find $m_t=173.2\pm 0.9\,$GeV~\cite{Lancaster:2011wr}.
\item {\it Higgs mass:}  Recent data from the CERN LHC~\cite{bib:CMS_SMHiggs,bib:ATLAS_SMHiggs} 
and Fermilab Tevatron~\cite{bib:Tevatron_SMHiggs} are consistent with {\it discovery of a Higgs boson} 
with $m_h= (125.5\pm 0.5)\ {\rm GeV}$ (combined results), while excluding a SM-like Higgs boson over 
a vast mass range around this value. In the SM, the Higgs mass is a free parameter, 
constrained only by unitarity bounds~\cite{Lee:1977eg}; in SUSY theories, 
quartic scalar terms are related to gauge couplings so that 
$m_h$ is constrained (within the MSSM, and including radiative corrections) 
to be $m_h\alt 135\,$GeV~\cite{Djouadi:2008gy}. 
The discovery of a light Higgs scalar with mass just below this bound lends credence to the MSSM 
as a viable effective field theory at the weak scale.
\item {\it Dark matter:} While none of the SM particles have the right properties to constitute
cold dark matter in the universe, SUSY theories offer several candidates~\cite{Steffen:2008qp}. 
These include the neutralino (a WIMP candidate), the gravitino or a singlet sneutrino. 
In SUSY theories where the strong $CP$
problem is solved via the Peccei-Quinn mechanism, there is the added possibility of mixed 
1. axion-neutralino~\cite{Choi:2008zq,Baer:2011hx,Baer:2011uz},
2. axion-axino~\cite{Rajagopal:1990yx,Covi:2001nw,Baer:2009ms} or 
3. axion-gravitino cold dark matter.
\item {\it Baryogenesis:} The measured baryon to photon ratio $\eta\simeq 10^{-10}$ is not possible to explain in
the SM. In SUSY theories, three prominent possibilities include 1. electroweak baryogenesis (now nearly excluded
by limits on $m_{\tst_1}$ and $m_h$~\cite{Curtin:2012aa}), 
2. thermal and non-thermal leptogenesis~\cite{Buchmuller:2005eh}, and 
3. Affleck-Dine baryo- or leptogenesis~\cite{Affleck:1984fy,Dine:1995kz}.
\end{itemize}

\subsection{Some problems for SUSY models}

While the above laundry list is certainly compelling for the existence of weak scale SUSY in nature,
we are faced with the fact that at present there is no evidence for direct superparticle production at high energy colliders, 
especially at the CERN Large Hadron Collider (LHC). The ATLAS and CMS experiments have accumulated $\sim 5\,$fb$^{-1}$
of integrated luminosity from $pp$ collisions at $\sqrt{s}=7\,$TeV in 2011 (LHC7), and 
$\sim 20\,$fb$^{-1}$ at $\sqrt{s}=8\,$TeV in 2012 (LHC8). 
Recent analyses by the CMS experiment~\cite{bib:CMS_alT} using $11.7\,$fb$^{-1}$ of data at 8 TeV have now excluded $m_{\tg}\alt 1500\,$GeV 
in the mSUGRA (also known as CMSSM) model
for the case of $m_{\tq}\simeq m_{\tg}$, while values of $m_{\tg}\alt 1000\,$GeV are excluded in the case where $m_{\tq}\gg m_{\tg}$.
Indeed, as recently as 2010~\cite{Bechtle:2009ty, Buchmueller:2010ai}, fits of the mSUGRA model to a variety of observables including EWPO, 
$(g-2)_\mu$, $B$-meson decay branching fractions and neutralino cold dark matter density 
predicted SUSY to lie exactly in this excluded range.
In addition, if the Higgs boson at $\sim 125\,$GeV turns out to be the light $CP$-even SUSY Higgs, 
then the minimal versions of gauge-mediated and anomaly-mediated SUSY breaking models will 
likely be ruled out~\cite{Arbey:2011ab}, since it is difficult to obtain such 
large values of $m_h$ in these models unless the sparticle mass spectra exist with a lightest MSSM particle of 
mass greater than about 5 TeV~\cite{Baer:2012uy}.

While the above results may seem disconcerting, at the same time they were not unanticipated by many theorists.
Whereas SUSY theories solve a host of problems as mentioned above, they also bring with them considerable
phenomenological baggage~\cite{Dienes:1997nq}. Some of these SUSY problems include the following:
\begin{itemize}
\item The SUSY flavor problem~\cite{Gabbiani:1996hi}:
In SUSY models based on gravity-mediation, it is generally expected that
large flavor-violating terms will occur in the Lagrangian~\cite{Kaplunovsky:1993rd}, giving rise to large contributions to the
kaon mass difference, and flavor violating decays such as $b\to s\gamma$ or $\mu\to e\gamma$.
Solutions to the SUSY flavor problem include 
1. degeneracy of matter scalar masses, in which case a SUSY GIM mechanism suppresses flavor violation~\cite{Dimopoulos:1981zb}, 
2. alignment of squark and quark mass matrices~\cite{Nir:1993mx}, or 
3. decoupling mainly of first/second generation scalars 
($m_{\tq,\tell}\agt 5-50\,$TeV)~\cite{Dine:1990jd,Cohen:1996vb,ArkaniHamed:1997ab}.\footnote{
Some degree of alignment or degeneracy would still be needed for the lower portion of this 
mass range.} 
Indeed, the SUSY flavor problem provided strong impetus
for the development of GMSB and AMSB models, where universality of scalars with the same quantum numbers is automatically expected.

\item The SUSY $CP$ problem: In this case, it is expected in gravity mediation that $CP$-violating phases in the
soft SUSY breaking terms and perhaps $\mu$ parameter will give rise to large electron and neutron (and other) electric dipole moments
(EDMs).
Solutions include dialing the $CP$-violating phases to zero, or decoupling with first generation scalars beyond the few TeV level. 
Models such as GMSB and AMSB are again not expected to have complex, $CP$-violating soft terms.

\item Proton decay in SUSY GUT theories: In SUSY GUT theories, the proton is expected to decay
to $K^+\bar{\nu}$ via colored Higgsino $\tilde h$ exchange. The lifetime is expected
to occur at levels below experimental limits~\cite{Murayama:2001ur}. 
Since $\Gamma_p\sim m_p^5/m_{\tilde h}^2m_{\tq}^2$, large squark masses can again suppress proton decay.

\item The gravitino problem~\cite{Weinberg:1982zq}: 
In models of gravity-mediation, the superhiggs mechanism generates SUSY breaking
by giving the gravitino a mass $m_{3/2}$. The gravitino mass sets the scale for the visible sector soft breaking terms,
and so one expects sparticle masses of order $m_{3/2}$. However, thermal production of gravitinos in the early universe
can lead to either 1. an overproduction of dark matter (here, the gravitinos would decay to the stable LSP, or even be the LSP),  
or 2. late-time decays of gravitinos at time scales $\agt 1\,$s after the Big Bang would lead to dissolution
of the light nuclei built up during Big Bang nucleosynthesis (BBN). Solutions to the gravitino problem include
1. a rather low re-heat temperature $T_R\alt 10^5\,$GeV after inflation so that thermal gravitino production is suppressed~\cite{Khlopov:1984pf} 
(but such low $T_R$ values conflict with some baryogenesis mechanisms such as leptogenesis, which 
seems to require $T_R\agt 10^9\,$GeV), 
2. a rather light gravitino with $m_{3/2}\ll 1\,$GeV, which enhances the goldstino coupling, or
3. a rather heavy gravitino $m_{3/2}\agt 5\,$TeV, which lowers the gravitino lifetime so that $\tau_{3/2}\alt 1$ sec, and
gravitinos decay before BBN~\cite{Kawasaki:2008qe}.
\end{itemize}

While some proposed solutions solve individual problems listed above ({\it e.g.} alignment for the SUSY flavor problem,
low $T_R$ for the gravitino problem, small phases for the SUSY $CP$ problem), there is one solution-- 
{\it decoupling of first/second generation matter scalars}-- which goes a long way to solving all 
four.\footnote{In gravity mediation, it is expected that the gravitino mass
$m_{3/2}$ sets the mass scale for the heaviest of the scalars; in this case, multi-TeV scalar masses would proceed from
a multi-TeV gravitino mass.} But what of fine-tuning constraints in SUSY models, which seemingly
require sparticle masses near the weak scale~\cite{Anderson:1994tr,Cassel:2009cx}?

\subsection{Fine-tuning in supersymmetric models}

As discussed previously, the most fundamental motivation for weak scale SUSY is that it 
provides a solution to the gauge hiearchy problem (GHP).
In SUSY models, the soft SUSY breaking parameters are intimately linked to the breakdown of electroweak
symmetry. This fact has motivated hope that the new SUSY matter states should not be too far removed
frpm the weak scale as typified by the $W$, $Z$ and $h$ masses, $\sim 100\,$GeV.
However, in the face of increasingly stringent LHC8 mass limits on superpartners, 
an awkward ``Little Hierarchy'' seems to be developing between the weak scale and the superpartner mass scale
which has led many to question whether the simple weak scale SUSY picture might be ruled out.
This problem is often referred to as the Litte Hierarchy Problem (LHP). 
In this subsection, we will review two approaches to quantify the severeness of the LHP, 
as manifested by possibly unnatural cancellations in building up the $Z$- or Higgs- boson masses. 

\subsubsection{High-scale fine-tuning measure $\Delta_{\mathrm{HS}}$: minimizing large logs}

In the SM, one may calculate the mass of the Higgs boson as
\be
m_h^2|_{\mathrm{phys}} =m_h^2|_{\mathrm{tree}}+\delta m_h^2|_{\mathrm{rad}}
\ee 
where $\delta m_h^2|_{\mathrm{rad}}=\frac{c}{16\pi^2}\Lambda^2$ and where $\Lambda$ represents the cutoff of 
quadratically divergent loop diagrams, which provides an upper limit to which the SM is considered a valid
effective field theory. The coefficient $c$ depends on the various SM couplings 
entering particular loop diagrams and here will be taken as $c\simeq 1$.
We may define a fine-tuning measure
\be
\Delta_{\mathrm{SM}}\equiv \delta m_h^2|_{\mathrm{rad}}/(m_h^2/2)
\ee
which compares the radiative correction to the physical Higgs boson mass.
Requiring $\Delta_{\mathrm{SM}} \alt 1$ then requires $\Lambda \simeq 1\,$TeV: {\it i.e.}, the SM should only be valid
up to at most the TeV scale.

Analogous reasoning has been applied to supersymmetric models~\cite{Kitano:2006gv}. 
In the MSSM, 
\be
m_h^2 \simeq \mu^2 +m_{H_u}^2|_{\mathrm{tree}}+\delta m_{H_u}^2|_{\mathrm{rad}}
\ee 
where
\be
\delta m_{H_u}^2|_{\mathrm{rad}}\simeq -\frac{3f_t^2}{8\pi^2}(m_{Q_3}^2+m_{U_3}^2+A_t^2)\ln\left(\Lambda^2/M_{\mathrm{SUSY}}^2 \right)
\ee
and where $\Lambda$ is again the cutoff scale which-- inspired by gauge coupling unification-- 
may be taken as high as $M_{\mathrm{GUT}}\simeq 2\times 10^{16}\,$GeV or even the reduced Planck mass $M_P$ 
and where $M_{\mathrm{SUSY}}^2 \simeq m_{\tst_1}m_{\tst_2}$. 
One may again create a fine-tuning measure $\Delta_{\mathrm{KN}}\equiv \delta m_{H_u}^2/(m_h^2/2)$, 
following the work of Kitano and Nomura~\cite{Kitano:2005wc}.
Using this, it has been asserted that low electroweak fine-tuning (EWFT) requires rather 
light third generation squarks:
\be \sqrt{m_{\tst_1}^2+m_{\tst_2}^2} \alt 600 \ {\rm GeV}
\frac{\sin\beta}{\sqrt{1+R_t^2}}\left(\frac{\log\frac{\Lambda}{{\rm
TeV}}}{3}\right)^{-1/2}\left(\frac{\Delta_{\mathrm{KN}}}{5}\right)^{1/2}\;,
\label{eq:papucci}
\ee 
where $R_t= A_t/\sqrt{m_{\tst_1}^2+m_{\tst_2}^2}$.  Taking $\Delta =10$
({\it i.e.} $\Delta^{-1}=0.1$ or 10\% EWFT) and $\Lambda$ as low as 20~TeV corresponds 
to Natural SUSY (NS)~\cite{Kitano:2006gv,Brust:2011tb,Papucci:2011wy}:
\bi
\item $m_{\tst_i},\ m_{\tb_1}\alt 600\ {\rm GeV}$,
\item $m_{\tg}\alt 1.5-2\ {\rm TeV}$.
\ei
The last of these conditions arises because the gluino enters the top-squark radiative corrections
$\delta m_{\tst_i}^2\sim (2g_s^2/3\pi^2)m_{\tg}^2 \times \log\Lambda$.  Setting
the $\log$ to unity and requiring $\delta m_{\tst_i}^2<m_{\tst_i}^2$
then implies $m_{\tg}\alt 3m_{\tst_i}$, or $m_{\tg}\alt 1.5-2\,$GeV for $\Delta\alt 10$.  
Taking $\Lambda$ as high as $M_{\mathrm{GUT}}$ leads to even tighter constraints:
$m_{\tst_{1,2}},m_{\tb_1}\alt 200\,$GeV and $m_{\tg}\alt 600\,$GeV, the latter
almost certainly in violation of LHC sparticle search constraints. 
Since (degenerate) first/second generation
squarks and sleptons enter the Higgs potential only at the two loop level, 
these can be much heavier: beyond LHC reach and also possibly
heavy enough to provide a (partial) decoupling solution to the SUSY flavor and $CP$
problems~\cite{Dine:1990jd}.
The NS models in the post-LHC8 period  suffer from three phenomenological problems arising from 
the very light top and bottom squarks: 1. large SUSY contributions to $BF(b\to s\gamma )$, 
2. small radiative corrections to $m_h$, thus making it difficult to generate $m_h\simeq 125\,$GeV at least within the 
MSSM and 3. there is so far no sign of light stops or sbottoms despite intensive searches at LHC8.

To bring the KN fine-tuning measure into closer accord with the measure
described below, we redefine it in terms of $m_Z^2/2$ instead of in terms of $m_h^2/2$, so that
$\Delta_{\mathrm{HS}}\simeq \delta m_{H_u}^2/(m_Z^2/2)$, where HS stands for ``high-scale''.
The quantity $\Delta_{\mathrm{HS}}$ is a rather severe measure of EWFT in that it doesn't account for
possible correlations amongst HS parameters which may lead to built-in cancellations 
between $m_{H_u}^2(\Lambda )$ and $\delta m_{H_u}^2$. 
As an example, the boundary condition $m_{H_u}^2(\Lambda = M_{\mathrm{GUT}})=m_0$ -- as in the mSUGRA/CMSSM model --
leads to large cancellations in $m_{H_u}^2(M_{\mathrm{weak}})$ in what has become 
known as the focus point (FP) region~\cite{Feng:2013pwa}. In the FP region, it is argued that the reduced EWFT
in the $m_0$ direction allows for squark/slepton masses far in excess of the values expected from $\Delta_{\mathrm{HS}}$. 

The lesson is that we must remember that HS models such as mSUGRA, NUHM2 etc. are nothing more than 
effective field theories,
albeit ones that are valid up to energy scales $Q\lesssim M_{\mathrm{GUT}}$. In this class of models,
the soft parameters merely serve to parametrize our ignorance of their true origin within 
some more fundamental theory. In the ultimate theory where the soft SUSY breaking parameters 
(and other parameters) are derived quantities, then correlations between soft parameters 
may exist which allow for large cancellations between $m_{H_u}^2 (\Lambda )$ and $\delta m_{H_u}^2$~\cite{Baer:2013ava}; 
in this case, the weak scale value value of $m_{H_u}^2$ becomes more relevant to fine-tuning discussions.

\subsubsection{Weak-scale fine-tuning and the Little Hierarchy Problem}

A more conservative EWFT measure has been advocated in Ref's.~\cite{Baer:2012up,Baer:2012mv,Baer:2012cf}.
Minimization of the MSSM scalar potential, including radiative corrections, leads to the well-known
relation
\be 
\frac{m_Z^2}{2} = \frac{m_{H_d}^2+\Sigma_d^d - (m_{H_u}^2+\Sigma_u^u) \tan^2\beta}{\tan^2\beta -1} -\mu^2 .
\label{eq:mZsSig}
\ee 
Noting that all entries in Eq.~\ref{eq:mZsSig} are defined at the weak scale, 
the {\rm electroweak fine-tuning parameter} 
\be 
\Delta_{\mathrm{EW}} \equiv max_i \left|C_i\right|/(m_Z^2/2)\;, 
\ee 
may be constructed, where $C_{H_d}=m_{H_d}^2/(\tan^2\beta -1)$, $C_{H_u}=-m_{H_u}^2\tan^2\beta /(\tan^2\beta -1)$ and $C_\mu =-\mu^2$. 
Also, $C_{\Sigma_u^u(k)} =-\Sigma_u^u(k)\tan^2\beta /(\tan^2\beta -1)$ and $C_{\Sigma_d^d(k)}=\Sigma_d^d(k)/(\tan^2\beta -1)$, 
where $k$ labels the various loop contributions included in Eq.~\ref{eq:mZsSig}.
Complete one-loop expressions for the $\Sigma_u^u$ and $\Sigma_d^d$ using the Coleman-Weinberg 
effective potential approach are given in the Appendix of Ref.~\cite{Baer:2012cf}.

Thus, $\Delta_{\mathrm{EW}}$ measures the largest {\it weak scale} contribution to the $Z$ mass. 
Model parameter choices which lead to low values of $\Delta_{\mathrm{EW}}$ are those which would naturally
generate a value of $m_Z\simeq 91.2\,$GeV. If any ranges of model parameters provide low $\Delta_{\mathrm{EW}}$, 
then one answers the fundamental question of the LHP: how can it be that
$m_Z$ and $m_h\simeq 100\,$GeV while gluino and squark masses lie at the TeV or beyond energy scale?

In order to achieve low $\Delta_{\mathrm{EW}}$,  it is necessary that $|m_{H_u}^2|$, $\mu^2$ and $|\Sigma_u^u|$ 
all be nearby to $m_Z^2/2$ to within a factor of a few~\cite{Baer:2012up,Baer:2012cf}. 
This implies the following:
\begin{enumerate}
\item $|\mu |$ is favored to be in the $100-300\,$GeV range (the closer to $m_Z$ the better).
\item  $|m_{H_u}^2|_{\mathrm{weak}} \simeq (100-300)^2\,$GeV$^2$. 
Such a small value of $m_{H_u}(M_{\mathrm{weak}})$ only occurs in the FP
region of mSUGRA, {\it i.e.} at very large $m_0$. In non-universal Higgs models (NUHM2), $m_{H_u}^2$ 
can be driven radiatively to small negative values at any $m_0$ and $m_{1/2}$ values.
\item To minimize the largest of radiative corrections $\Sigma_u^u(\tst_{1,2})$, large stop mixing 
$A_0\simeq \pm 1.6 m_0$ is required. The large mixing both softens the top-squark radiative corrections
while raising $m_h$ up to the $\sim 125\,$GeV level.
\end{enumerate}

The measure $\Delta_{\mathrm{EW}}$ listed above is created from only weak scale MSSM parameters; it 
contains no information about any possible high scale origin of the soft parameters. 
In this sense, low $\Delta_{\mathrm{EW}}$ captures a {\it minimal}, non-negotiable EWFT required of even 
high scale SUSY models.

Constrained models such as mSUGRA, mGMSB and mAMSB have all been found to be highly fine-tuned under 
$\Delta_{\mathrm{EW}}$. 
However, models like NUHM2 allow for $\Delta_{\mathrm{EW}}$ as low as $5-10$ to be generated. 
For such cases, the Little Hierarchy is not a Problem: the necessary condition for low
EWFT using $\Delta_{\mathrm{EW}}$ is that only $|\mu |$, $m_{H_u}(M_{\mathrm{weak}})$ and the various $\Sigma_u^u$ 
need be close to the $m_Z,\ m_h$ scale.

The low $\Delta_{\mathrm{EW}}$ models are typified by the presence of light higgsinos 
$m_{\twpm_1},\ m_{\tz_{1,2}}\simeq 100-300\,$GeV.
Also, top squark masses can be significantly heavier than in NS models, with 
$m_{\tst_1}\simeq 1-2\,$TeV and $m_{\tst_2}\simeq 2-4\,$TeV. Likewise, gluino masses tend to be bounded by
about 5 TeV lest they contribute too much to uplifting the top squark masses. 
The spectrum of higgsinos is highly compressed amongst themselves: in models with gaugino mass
unification, the higgsino mass gaps are typically $\sim 10-30\,$GeV, since $|M_3|\lesssim 5\,$TeV requires
also upper limits on $M_1$ and $M_2$. If gaugino  mass unification is relaxed, then 
$M_1$ and $M_2$ can be heavier, leading to even smaller mass gaps as low as the GeV range.
But even for $\sim 10-30\,$GeV mass splittings, the higgsino decays give rise to very soft visible energy 
release which makes their detection at LHC very difficult. 
Also, since $m_{\tg}\simeq 1-5\,$TeV, $m_{\tst_1}\simeq 1-2\,$TeV and $m_{\tst_2}\simeq 2-4\,$TeV, these colored
SUSY particles may be too massive to be revealed in LHC SUSY searches.
However, the spectrum of light higgsinos should be easily visible to a linear $e^+e^-$ collider 
operating with $\sqrt{s}\agt 2|\mu |$.

\subsection{Remainder of this report}

The remainder of this report is geared towards presenting a new set of supersymmetry benchmark
models suitable for ILC investigations, while maintaining consistency with the latest 
indirect and direct constraints on supersymmetric models, especially taking into 
account what has been learned from recent LHC searches. In Sec.~\ref{sec:constraints}, we
briefly summarize current indirect constraints on SUSY models, and also discuss 
the current status of SUSY dark matter.
In Sec.~\ref{sec:lhc}, we present a summary of the most recent results from LHC searches for SUSY and
discovery of the Higgs boson. 
In Sec.~\ref{sec:BMs}, we present a variety of new post LHC8 benchmark points
for ILC studies. These new benchmarks reflect a movement away from previous studies within the
mSUGRA/CMSSM model. Some models have been selected due to their theoretical motivation
({\it e.g.} natural SUSY and its relatives), while others have been selected for their diversity of
phenomenology which may be expected at ILC.
In Sec.~\ref{sec:conclude}, we present a brief summary and outlook for physics prospects at the ILC.

\section{Indirect constraints on SUSY models}
\label{sec:constraints}

In this section, we review briefly indirect constraints on SUSY models from the measurement of the anomalous magnetic moment of the muon, rare $B$-decay branching fractions along with an updated discussion of the role of
dark matter in SUSY models.

\subsection{$(g-2)_\mu$ status}

The magnetic moment of the muon $a_\mu\equiv\frac{(g-2)_\mu}{2}$ 
was measured by the Muon $g-2$ Collaboration~\cite{Bennett:2006fi} 
and has been found to give a $3.6\sigma$ discrepancy with SM calculations based on $e^+e^-$ data~\cite{Davier:2010nc}:
$\Delta a_\mu =a_\mu^{\mathrm{meas}}-a_\mu^{\mathrm{SM}}[e^+e^-]=(28.7\pm 8.0)\times 10^{-10}$. When $\tau$-decay data are used
to estimate the hadronic vacuum polarization contribution rather than low energy $e^+e^-$ annihilation 
data, the discrepancy reduces to $2.4\sigma$ , corrensponding to 
$\Delta a_\mu =a_\mu^{\mathrm{meas}}-a_\mu^{\mathrm{SM}}[\tau]=(19.5\pm 8.3)\times 10^{-10}$.

The SUSY contribution to the muon magnetic moment is~\cite{Moroi:1995yh}
$\Delta a_\mu^{\mathrm{SUSY}}\sim \frac{m_\mu^2\mu M_i\tan\beta}{M_{\mathrm{SUSY}}^4}$ where $i=1,2$ 
stands for
electroweak gaugino masses and $M_{\mathrm{SUSY}}$ is the characteristic sparticle mass 
circulating in the
muon-muon-photon vertex correction: here, $m_{\tmu_{L,R}}$, $m_{\tnu_\mu}$, $m_{\twp_i}$ and $m_{\tz_j}$. 
Attempts to explain the muon $g-2$ anomaly using supersymmetry usually invoke sparticle mass 
spectra with relatively light smuons and/or large $\tan\beta$ 
(see {\it e.g.} Ref.~\cite{Feng:2001tr}). 
Some SUSY models where $m_{\tmu_{L,R}}$ is correlated with squark masses (such as mSUGRA) 
are now highly stressed to explain the $(g-2)_\mu$ anomaly. In addition, since naturalness favors a low value
of $|\mu |$, tension again arises between a large contribution to $\Delta a_\mu^{\mathrm{SUSY}}$ and naturalness conditions.
These tensions motivate scenarios with non-universal scalar masses. Of the benchmark scenarios discussed in the 
following, some feature light smuons which raise $(g-2)_\mu$  to its experimental value, while others are 
compatible with the Standard Model prediction.

\subsection{$b\to s\gamma$}

The combination of several measurements of the $b\to s\gamma $ branching fraction by the 
Heavy Flavor Averaging Group (HFAG)~\cite{Amhis:2012bh}
finds that $BF(b\to s\gamma )=(3.55\pm 0.26)\times 10^{-4}$.
This is somewhat higher than the SM prediction~\cite{Misiak:2006zs} of 
$BF^{\mathrm{SM}}(b\to s\gamma )=(3.15\pm 0.23)\times 10^{-4}$. SUSY contributions to the
$b\to s\gamma$ decay rate come mainly from chargino-top-squark loops and
loops containing charged Higgs bosons, and so are large when these particles are light 
and when $\tan\beta$ is large~\cite{Baer:1996kv}. Most SUSY model predictions for
$BF(b\to s\gamma )$ decay assume minimal flavor violation (MFV); if this assumption is relaxed, then
additional flavor-sector contributions to $BF(b\to s\gamma )$ can occur.

\subsection{$B_s\to \mu^+\mu^-$}

Recently, the LHCb collaboration has discovered an excess over the
background for the decay $B_s\to\mu^+\mu^-$~\cite{lhcb}!  They find a
branching fraction of $BF(B_s\to\mu^+\mu^- )=3.2^{+1.5}_{-1.2}\times 10^{-9}$ 
in accord with the SM prediction of $(3.2\pm 0.2)\times 10^{-9}$~\cite{bmm_sm}. 
In supersymmetric models, this flavor-changing decay occurs
through pseudoscalar Higgs $A$ exchange~\cite{Babu:1999hn,Mizukoshi:2002gs}, and the
contribution to the branching fraction from SUSY is proportional to
$\frac{(\tan\beta)^6}{m_A^4}$.  
Thus, the decay is most constraining at large $\tan\beta\simeq 50$ (as occurs in
Yukawa-unified models) and at low $m_A\simeq 100-200\,$GeV. 

\subsection{$B_u\to\tau^+\nu_\tau$}

The branching fraction for $B_u\to\tau^+\nu_\tau$ decay is calculated~\cite{Eriksson:2008cx} in the SM to be
$BF(B_u\to\tau^+\nu_\tau )=(1.10\pm 0.29)\times 10^{-4}$. This is to be compared to the value from
the HFAG~\cite{Amhis:2012bh}, which finds a measured value of 
$BF(B_u\to\tau^+\nu_\tau )=(1.67\pm 0.3)\times 10^{-4}$, somewhat beyond -- but not disagreeing with -- the SM prediction.
The main contribution from SUSY arises due to tree-level charged Higgs exchange, 
and is large at large $\tan\beta$ and low $m_{H^+}$.

\subsection{Dark matter}

During the past several decades, a very compelling and simple scenario has
emerged to explain the presence of dark matter in the universe with an abundance roughly
five times that of baryonic matter. The WIMP miracle scenario posits that 
weakly interacting massive particles would be in thermal equilibrium with the cosmic
plasma at very high temperatures $T\agt m_{\mathrm{WIMP}}$. As the universe expands and cools, 
the WIMP particles would freeze out of thermal equilibrium, locking in a relic abundance
that depends inversely on the thermally-averaged WIMP (co)-annihilation 
cross section~\cite{Lee:1977ua}.
The WIMP ``miracle'' occurs in that a weak strength annihilation cross section gives
roughly the measured relic abundance provided the WIMP mass is of the order of the 
weak scale~\cite{Baltz:2006fm}.
The lightest neutralino of SUSY models has been touted as a 
protypical WIMP candidate~\cite{Goldberg:1983nd,Ellis:1983ew,Jungman:1995df}.

While the WIMP miracle scenario is both simple and engaging, it is now clear that
it suffers from several problems in the case of SUSY theories. 
The first of these is that in general SUSY theories where the lightest neutralino
plays the role of a thermally produced WIMP, the calculated relic abundance $\Omega_{\chi}h^2$
is in fact typically  $2-4$ orders of magnitude larger than the measured abundance
$\Omega_{\mathrm{CDM}}^{\mathrm{meas}}h^2 = 0.115\pm 0.002$~\cite{Bennett:2012fp,Ade:2013zuv} 
in the case of a bino-like neutralino, and $1-2$
orders of magnitude lower than measurements in the case of wino- or higgsino-like
neutralinos~\cite{Baer:2010wm}. In fact, rather strong co-annihilation, resonance annihilation or
mixed bino-higgsino or mixed wino-bino annihilation is needed to obtain the measured
dark matter abundance. Each of these scenarios typically requires considerable large 
fine-tuning of parameters to gain the measured dark matter abundance~\cite{Baer:2009vr}. 
The case where neutralinos naturally give the measured CDM
abundance is when one has a bino-like neutralino annihilating via slepton exchange
with slepton masses in the $50-70\,$GeV range: such mass values were long ago ruled out by slepton
searches at LEP2~\cite{bib:LEP2sleptons}.

The second problem with the SUSY WIMP miracle scenario is that it neglects the gravitino, which
is an essential component of theories based on supergravity. Gravitinos can be produced 
thermally at high rates at high re-heat temperatures $T_R$ after inflation. 
If $m_{\tG}>m_{\mathrm{LSP}}$, then gravitino decays into a stable LSP can overproduce 
dark matter for $T_R\agt 10^{10}\,$GeV. 
Even at much lower $T_R\simeq 10^5-10^{10}\,$GeV, 
thermal production of gravitinos followed by late decays 
(since gravitino decays are suppressed by the Planck scale) tend to dissociate light nuclei
produced in the early universe, thus destroying the successful picture of 
Big Bang nucleosynthesis~\cite{Kawasaki:2008qe}.

The third problem is that the SUSY WIMP scenario neglects at least two very compelling 
new physics effects that would have a strong influence on dark matter production in the 
early universe. 
\bi
\item The first of these is that string theory seems to require 
the presence of at least one light ($\sim 10-100\,$TeV) moduli field~\cite{Acharya:2010af}. 
The moduli can be produced at large rates in the early universe and decay 
at times $\sim 10^{-1}-10^5\,$s after the Big Bang. 
Depending on their branching fractions, they could
either feed additional LSPs into the cosmic plasma~\cite{Moroi:1999zb}, 
or decay mainly to SM particles, thus diluting all relics present at the time of 
decay~\cite{Gelmini:2006pq}.
\item The second neglected effect is the strong $CP$ problem, which is deeply rooted in QCD
phenomenology~\cite{Peccei:2006as}. 
After more than three decades, the most compelling solution to the strong $CP$
problem is the hypothesis of a Peccei-Quinn axial symmetry whose breaking gives rise to
axion particles with mass $\sim 10^{-6}-10^{-9}\,$eV~\cite{Kim:2008hd}. The axions can be produced non-thermally
via coherent oscillations (CO)~\cite{Abbott:1982af,Preskill:1982cy,Dine:1982ah}, 
and also would constitute a portion of the dark matter. 
In SUSY theories, the axions are accompanied by $R$-odd spin-${1\over 2}$ axinos $\ta$ and
$R$-even spin-0 saxions $s$~\cite{Nilles:1981py}. 
Thermal production of axinos and CO-production of saxions can either
feed more dark matter particles into the cosmic plasma, or inject additional entropy, thus diluting
all relics present at the time of decay. Theoretical predictions for the relic abundance of
dark matter in these scenarios are available but very model-dependent. In the case of 
mixed axion-neutralino dark matter, it is usually very difficult to lower a standard
overabundance of neutralinos, but it is also very easy to bolster a standard underabundance {\it e.g.}
by decay-produced neutralino reannihilation at temperatures below standard 
freeze-out~\cite{Baer:2011uz,Bae:2013qr}.
This latter case may lead one to consider SUSY models with a standard underabundance
of wino-like or higgsino-like neutralinos as perhaps the more compelling possibility for
CDM. In the case of mixed axion-neutralino CDM, it can be very model-dependent
whether the axion or the neutralino dominates the DM abundance, and  cases where
there is a comparable admixture of both are possible. 
\ei
The upshot for ILC or LHC physics is that one shouldn't take dark matter abundance constraints on
SUSY theories too seriously at this point in time.

\subsubsection{Status of WIMP dark matter searches}

As of spring 2013, a variety of direct and indirect WIMP dark matter detection searches
are ongoing. Several experiments -- DAMA/Libra, CoGent, Cresst and CDMS -- claim excess signal
rates beyond expected backgrounds. These various excesses can be interpreted in terms of
a 5-10 GeV WIMP particle, although the four results seem at first sight inconsistent 
with each other. It is also possible that muon- or nuclear-decay induced 
neutron backgrounds -- which are very difficult to estimate -- contribute to the 
excesses. Numerous theoretical and experimental analyses are ongoing to sort 
the situation out. A  WIMP particle of several GeV seems hard to accommodate in SUSY theories
(but see {\it e.g.} Ref.~\cite{Bottino:2003cz}). 

There also exist excesses of positrons in cosmic rays above expected backgrounds, first
observed by the Pamela collaboration~\cite{Adriani:2008zr}, 
and later by the Fermi-LAT~\cite{FermiLAT:2011ab} and AMS~\cite{Aguilar:2013qda} experiments.
While this excess could be understood in terms of very massive
WIMPs of order hundreds of GeV, it is unclear at present whether the positrons arise from exotic
astrophysical sources such as pulsars~\cite{Profumo:2008ms,Barger:2009yt} 
or simply from rare mis-identification of cosmic protons.
A further possible indirect WIMP signal is the $130\,$GeV gamma ray line seen in 
some portions of Fermi data~\cite{Weniger:2012tx}.
This could be interpreted as $\chi\chi\to\gamma\gamma$ which occurs via a box-diagram in SUSY.
While intriguing, this signal seems incompatible with SUSY in that one would also expect a much 
larger continuum
distribution of photons from direct WIMP annihilation, which doesn't seem to appear.

A variety of other direct WIMP search experiments have probed deeply into WIMP-model
parameter space, with no apparent excesses above SM background. At this time, the best limits come
from the XENON100 experiment~\cite{Aprile:2012nq}, 
which excludes WIMP-proton scattering cross sections of
$\sigma (\chi p)\agt 2\times 10^{-9}\,$pb at 90\%CL for $m_{\mathrm{WIMP}}\simeq 100\,$GeV.
The XENON100, LUX and CDMS experiments seem poised to decisively probe the SUSY parameter space
associated with well-tempered (mixed bino-higgsino) dark matter~\cite{Baer:2006te,Feng:2010ef} 
(as occurs for instance in focus point SUSY of the mSUGRA model) in the current round of data taking.

\subsubsection{Gravitino dark matter}

It is possible in SUSY theories that gravitinos are the lightest SUSY particle, 
and could fill the role of dark matter. In gravity-mediation, the gravitino is expected 
to have mass of order the weak scale. In this case, late decays of thermally produced
neutralinos into gravitinos are often in conflict with BBN constraints. If the gravitinos are
much lighter, well below the GeV scale, then their goldstino coupling is enhanced
and BBN constraints can be evaded. This scenario tends to occur for instance in gauge-mediated SUSY theories. 
The simplest GMSB scenarios now appear in
conflict with the LHC discovery of a Higgs boson with $m_h\simeq 125\,$GeV~\cite{Arbey:2011ab,Baer:2012uy}. 
We will, however, present an example of a non-minimal GMSB 
model which is compatible with the Higgs mass measurement.

\section{LHC results}
\label{sec:lhc}

In this section, we present a very brief summary of the status of 
LHC searches for SUSY Higgs bosons and for SUSY particles as of mid 2013.

\subsection{Impact of Higgs searches}

\subsubsection{SM-like Higgs scalar}

%
After the discovery of a new boson in summer 2012 by the ATLAS~\cite{bib:ATLAS_SMHiggs} and CMS experiments~\cite{bib:CMS_SMHiggs}, the analysis of the full LHC7 and LHC8 data sets showed that
it is indeed a Higgs boson with roughly SM-like signal strengths in various channels~\cite{bib:ATLAS_mass_mu,bib:CMS_mass_mu}. Its mass has been determined to $125.7\pm0.3$(stat.)$\pm0.3$(syst.)~GeV  by CMS, while ATLAS finds $125.5\pm0.2$(stat.)$+0.5-0.6$(syst.)~GeV. In addition the Tevatron experiments reported a $3\sigma$ excess in  the search for the SM Higgs boson in the $W/Z+ b\bar{b}$ channel~\cite{bib:Tevatron_SMHiggs}.

Although the observed Higgs boson looks SM-like, the signal strengths in the various channels have currently still large uncertainties of $30$ to $100\%$ and therefore easily leave room for deviations of up to $10$ or $20\%$ as they typically 
appear in the MSSM.

\subsubsection{Non-standard Higgs bosons}

Searches by ATLAS and CMS for $H,\ A\to\tau^+\tau^-$ now exclude a large portion of the
$m_A\ vs.\ \tan\beta$ plane~\cite{bib:CMS_SUSYHiggs_tautau,bib:ATLAS_SUSYHiggs_tautau}. 
In particular, the region around $\tan\beta\simeq 50$, 
which is favored by Yukawa-unified SUSY GUT theories, now excludes $m_A<500\,$GeV.
For $\tan\beta =10$, the range $120\ {\rm GeV}<m_A<220\,$GeV is excluded in the $m_h^{\mathrm{max}}$ scenario with $M_{\mathrm{SUSY}} = 1\,$TeV.
ATLAS and CMS also searched for charged Higgs bosons produced in decays of top quarks.
Both experiments exclude charged Higgs masses between $90$ and about $150\,$GeV for $\tan{\beta} \simeq 20$ (and in the 
case of ATLAS also for $\tan{\beta}$ below $4$) 
in the $m_h^{\mathrm{max}}$ scenario with $M_{\mathrm{SUSY}} = 1\,$TeV~\cite{bib:ATLAS_SUSYHiggs_taunu, bib:CMS_SUSYHiggs_taunu}. 
For $\tan{\beta} \simeq 10$, no charged Higgs mass is excluded beyond the LEP limit of $80.0\,$GeV~\cite{Abbiendi:2013hk}.

\subsubsection{Impact of Higgs searches on SUSY models}

\begin{figure}[htb]
  \begin{center}
\includegraphics[width=0.45\textwidth]{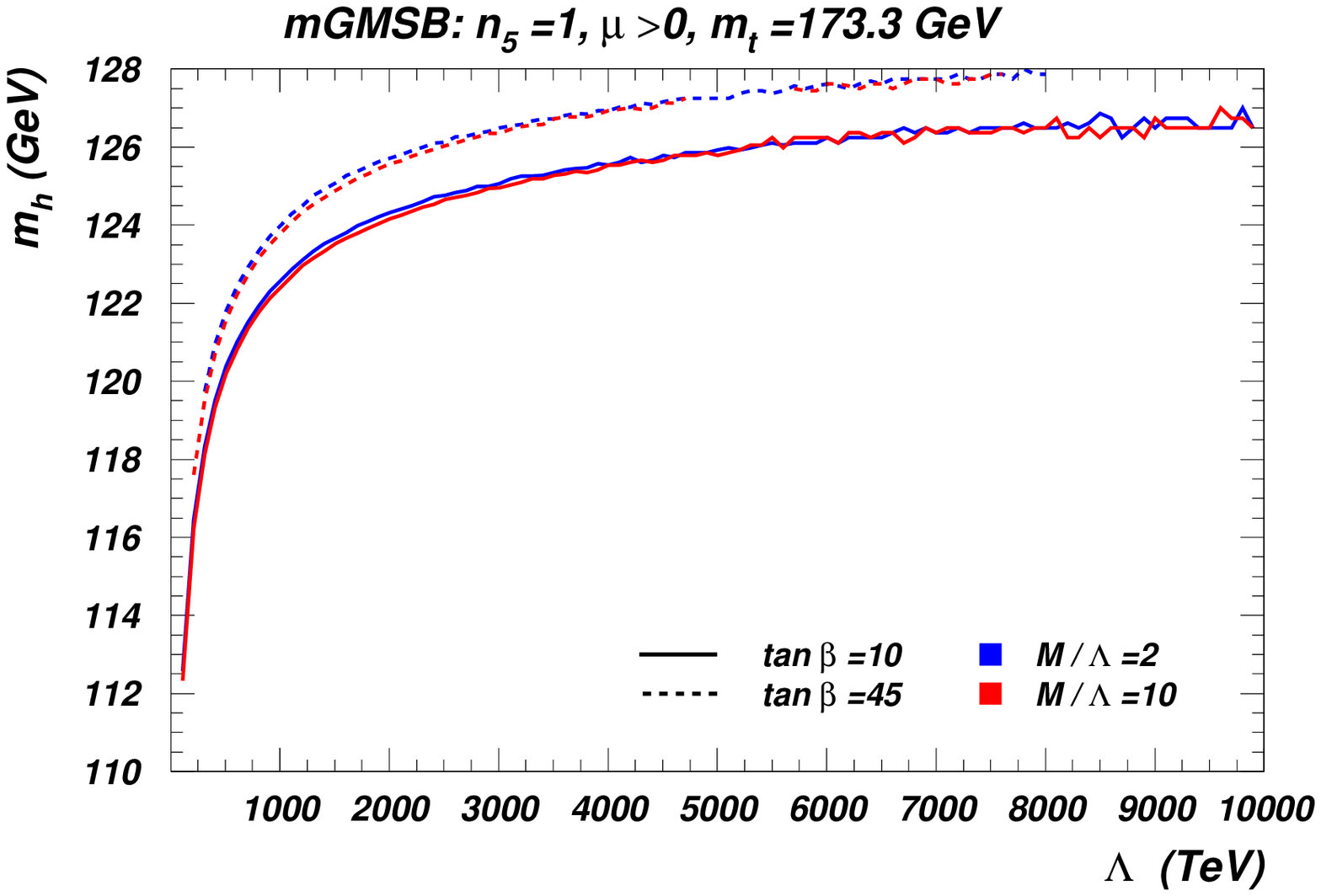}
\hspace{0.1cm}
\includegraphics[width=0.45\textwidth]{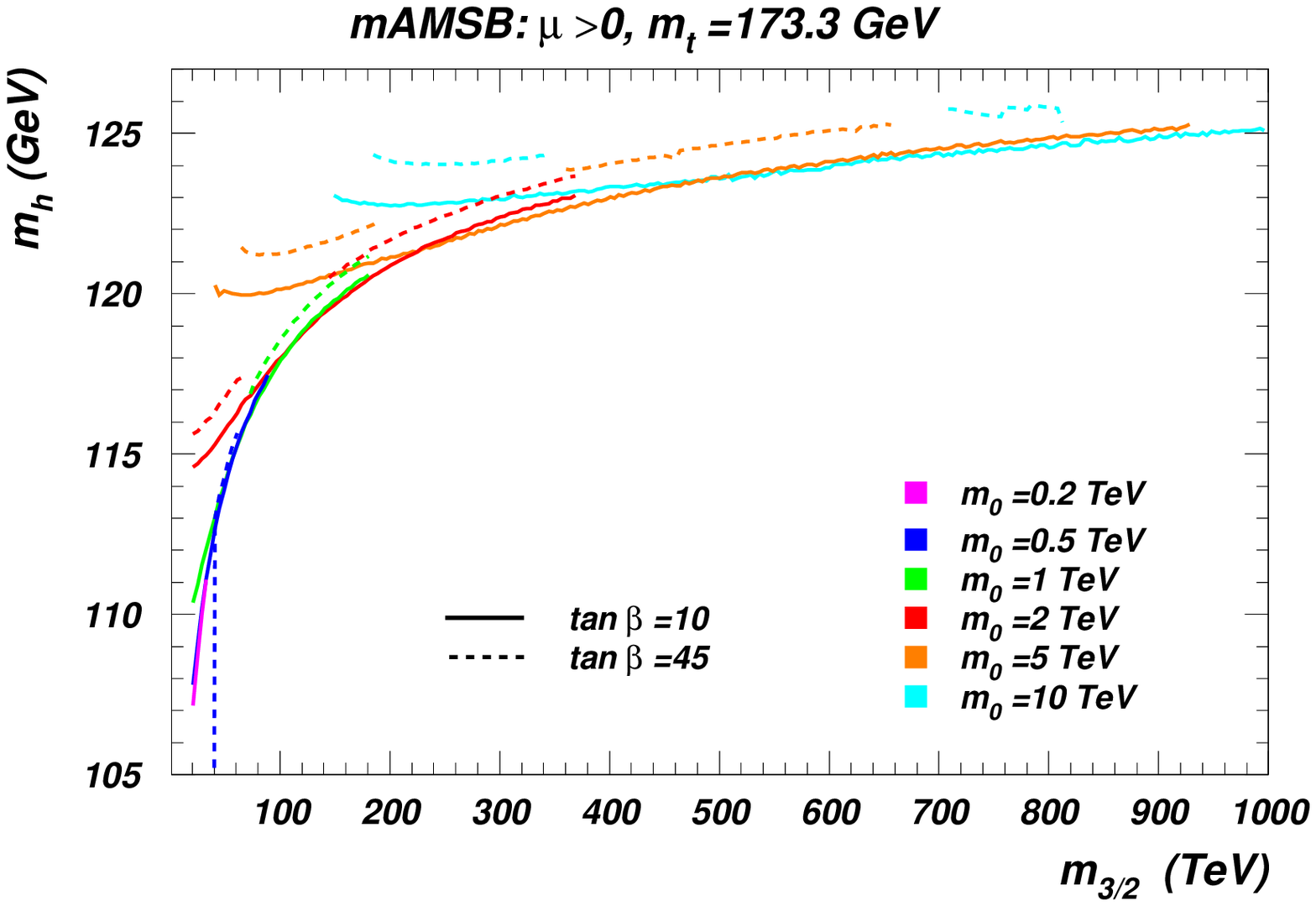}
  \end{center}
  \caption{Value of $m_h$ in mGMSB and in mAMSB versus $\Lambda$ and $m_{3/2}$
from~\cite{Baer:2012uy}.}
\label{fig:Mh_GMSB_AMSB}
\end{figure}

A Higgs mass of $m_h \simeq 125\,$GeV lies below the value of $m_h\simeq 135\,$GeV which is allowed
by calculations within the MSSM.
However, such a large value of $m_h$ requires large radiative corrections and 
large mixing in the top squark sector. In models such as mSUGRA, trilinear soft parameters $A_0\simeq \pm 2m_0$
are thus preferred, and values of $A_0\simeq 0$ would be ruled out~\cite{Baer:2011ab,Heinemeyer:2011aa}.
In other constrained models such as the minimal versions of GMSB or AMSB,
Higgs masses of $125\,$GeV require even the lightest of sparticles to be in the multi-TeV range~\cite{Baer:2012uy}, 
as illustrated in Figure~\ref{fig:Mh_GMSB_AMSB}.
%

In the mSUGRA/CMSSM model, requiring a Higgs mass of about $125\,$GeV pushes the best fit point in
$m_0$ and $m_{\frac{1}{2}}$ space into the multi-TeV range~\cite{Baer:2011ab} and makes global fits 
of the model to data increasingly difficult~\cite{Bechtle:2012zk,Buchmueller:2012hv}. This has provided motivation
for extending the MSSM with gauge singlets~\cite{Hall:2011aa,King:2012is} or vector-like matter~\cite{Martin:2009bg} 
both of which allow for somewhat heavier values of $m_h$.

 While the experimental uncertainty of the mass has shrunken to about $0.5\,$GeV, a considerable theoretical uncertainty
needs to be taken into account when comparing this number to Higgs mass predictions calculated from SUSY parameters.
We therefore consider Higgs masses in the range of $122-128\,$GeV to be in agreement with current observations.

Although the interpretation of the $125\,$GeV particle as the light, $CP$-even Higgs boson $h$
is the most obvious one, it is not the only possibility. With the current precision on the 
signal strengths and the current limits on the heavy Higgs bosons and SUSY particles, the 
heavy $CP$-even Higgs boson, $H$, could be SM-like and the one observed at the 
LHC~\cite{Heinemeyer:2011aa, Benbrik:2012rm, Bechtle:2012jw, Drees:2012fb}. 

\subsection{Review of sparticle searches at LHC}

\subsubsection{Gluinos and first/second generation squarks}

The ATLAS and CMS collaborations have searched for multi-jet$+\eslt$ events 
arising from gluino and squark pair production in 20 fb$^{-1}$ of data taken 
at $\sqrt{s}=8\,$TeV~\cite{bib:ATLAS_jets, Chatrchyan:2013lya}. 
In a simplified squark-gluino-LSP model, they exclude up to $m_{\tg}\alt 1.4\,$TeV in the limit of very heavy squark masses, 
while $m_{\tg}\alt 1.7\,$TeV is excluded for  $m_{\tq}\simeq m_{\tg}$. Here, $m_{\tq}$ refers to a 
generic first generation squark mass scale, since these are the ones whose production rates 
depend strongly on valence quark PDFs in the proton. 

If the gluino decays dominantly via third generation squarks, the gluino mass limits are somewhat
weaker, typically in the range of $1.0$ to $1.2\,$TeV, again depending on the exact decay 
chain~\cite{bib:ATLAS_multijets, bib:ATLAS_dileptonbjets, Chatrchyan:2013wxa, Chatrchyan:2012paa, bib:CMS_leptonbjets, bib:CMS_trileptonbjets}. Similar limits have been 
found for the case of intermediate charginos~\cite{bib:ATLAS_jets}.

It has been shown that these limits get considerable weaker if not all squarks
are mass degenerate or in case of compressed spectra. In the latter case, the best sensitivity is
often obtained from mono-jet searches~\cite{Dreiner:2012gx}, and limits on squark masses can reduce to as low 
as $340\,$GeV.

%

\subsubsection{Sbottom and Stop}

Motivated by naturalness, ATLAS and CMS recently put a lot of emphasis on the search for direct production of third generation squarks. They searched for top squarks decaying to  $t \tz_1$~\cite{bib:ATLAS_3gen_stop-0lepton, bib:ATLAS_3gen_stop-1lepton, bib:CMS_stop_lepton}, as well as for $\tst_1 \ra b W \tz_1$~\cite{bib:ATLAS_3gen_stop_2lepton}, $\tst_1 \ra b \twp_1$~\cite{bib:ATLAS_3gen_stop-1lepton, bib:ATLAS_3gen_stop_2lepton, bib:ATLAS_3gen_2bjets, bib:CMS_stop_lepton} and $\tb_1 \to b Z \tz_1 / t W \tz_1$~\cite{bib:CMS_sbottom}. In the easiest case, namely for mass differences so large that $m_{\tst_1} > m_t + m_{\tz_1}$, stop masses up to about $700\,$GeV and $\tz_1$ masses up to $200\,$GeV have been probed. However the resulting exclusions leave substantial uncovered territory at lower stop masses, where especially the regions near 
$m_{\tst_1} = m_t + m_{\tz_1}$ and $m_{\tst_1} = m_b + m_W + m_{\tz_1}$ are difficult. The exclusions for $\tb_1 \to b \tz_1$ also apply to top squark pair production in case of $\tst_1\to b\twp_1$ and
the $\twpm_1$ decays to soft, nearly invisible particles, as would be expected in natural SUSY due to the small mass difference between the higgsinos. However, these exclusions  depend strongly on the assumptions for the chargino and LSP masses. 
For instance, for $m_{\twpm_1} = 150\,$GeV and $m_{\tz_1}$ between about $80$ and $140\,$GeV, no stop mass is excluded for the $\tst_1\to b\twp_1$ decay. Stop-LSP mass differences smaller than approximately $40\,$GeV have not been probed at all sofar.

In the context of GMSB with the $\tz_1$ as higgsino-like NLSP and a gravitino $\tG$ LSP, 
ATLAS searched for direct top squark pair production, followed by $\tst_1\to b\twp_1$ or, 
when kinematically allowed, also $t\tz_1$. 
Based on $2$~fb$^{-1}$, they probe top squark masses up to $600\,$GeV~\cite{bib:ATLAS_GMSBstop}. 
This limit relies on the GMSB specific decay of the $\tz_1$ into $Z \tG$, especially on two (same flavour, opposite sign) leptons 
consistent with the $Z$ mass.

\subsubsection{Electroweakinos}

Direct production of neutralinos, charginos and sleptons does not rely on coloured SUSY partners to be within reach.
However the cross-sections at the LHC are significantly lower than for squark and gluino production, so that the obtained limits are considerably weaker. The dominant electroweak production mechanism at the LHC is $\tz_2 \twp_1$ production, which has been searched for by ATLAS based on the full $8\,$TeV dataset~\cite{bib:ATLAS_ewkino_2lepton, bib:ATLAS_ewkino_3lepton, bib:ATLAS_ewkino_taus}, and by CMS based on roughly half the $8\,$TeV data~\cite{bib:CMS_ewkino}.

The sensitivity of LHC searches for this mode depend stongly on the details of the SUSY spectrum, so that the strongest loophole-free limits on the chargino mass are still the limits of $m_{\twpm_1} > 103.5\,$GeV for mass differences larger than $3\,$GeV and $m_{\twpm_1} > 92.4\,$GeV for smaller mass differences down to $60\,$MeV obtained from LEP data~\cite{bib:LEP_chargino}.  All LHC searches assume that $\tz_2$ and $\twpm_1$ are mass-degenerate, but have a sizable mass difference of at least $40$ to $60\,$GeV to the $\tz_1$, which is the typical pattern in the case of a bino-like LSP. The strongest limits are obtained when assuming that the lighter set of sleptons, 
including the third generation, is mass degenerate and fulfills $m_{\tl}=(m_{\twpm_1}-m_{\tz_1})/2$, 
which maximizes the lepton momenta and thus the acceptance. In this case, chargino masses from $100$ to $650\,$GeV are excluded for $m_{\tz_1} = 60\,$GeV. For higher LSP masses, the limit does not reach down to the LEP limit. CMS also studied the case where the slepton masses are either at $5\%$ or $95\%$ of the chargino-LSP mass difference, which leads to a weakening of the limits due to less favourable lepton momentum distributions. 

Both collaborations also studied the case of only the $\ttau$ (and in case of ATLAS the $\tau$-sneutrino) appearing in the decay chains with a mass halfway between chargino and LSP masses~\cite{bib:CMS_ewkino, bib:ATLAS_ewkino_taus}, and $\te$ and $\tmu$ being more heavy. In the most optimistic case, {\it i.e.} for $m_{\ttau}=m_{\tnu_{\tau}}=(m_{\twpm_1}-m_{\tz_1})/2$, chargino masses up to $350\,$GeV can be probed for a massless LSP. For $m_{\tz_1} = 100\,$GeV, no chargino masses are excluded beyond the LEP limit. This means that in the particularly interesting case of a small $\ttau$-$\tz_1$ mass difference, as is 
{\it e.g.} required to obtain a sufficiently low dark matter relic density via $\ttau$-coannihilation, the chargino mass is not constrained beyond
the LEP results.

Finally, also the possibility that all sleptons are heavier than the chargino has been studied. 
In this case, chargino and neutralino decay via real or virtual $W$ and $Z$ (or Higgs) bosons, 
depending on the mass differences.
ATLAS~\cite{bib:ATLAS_ewkino_3lepton} and CMS~\cite{bib:CMS_ewkino} exclude chargino masses up to about $300$ to $350$ for LSP masses below $70\,$GeV. Above $m_{\tz_1} = 120\,$GeV, no chargino masses have been excluded.

\subsubsection{Sleptons}

Recently, ATLAS and CMS obtained also first results on slepton pair 
production~\cite{bib:ATLAS_ewkino_2lepton, bib:CMS_ewkino}. 
For LSP masses below $30\,$GeV, they reach down to the LEP limits and extend up to $300\,$GeV for left-handed and up to $230\,$GeV for right-handed selectons and smuons. For higher LSP masses, the LHC exclusions do not connect to the LEP limit, and leave an untouched corridor corresponding to mass differences to the LSP below about $70\,$GeV. It should be noted that this untouched corridor is in the only region where the conditions of this simplified model can be realised without giving up gaugino mass unification at the GUT-scale. No slepton masses are excluded for LSP masses above $90\,$GeV ($150\,$GeV) in the case of right-handed (left-handed) sleptons. 

\section{Implications for ILC and benchmark points}
\label{sec:BMs}

The results from the previous sections, when summarized, yield the following grand picture:

\begin{itemize}
\item{\bf Squarks and gluinos:} Ironically, the strongest LHC limits on sparticle masses apply to 
the first generation squarks and gluinos, while these are the most remotely connected to the determination of 
the electroweak scale, and to the weak boson masses. So while $m_{\tg}\agt 1.5\,$TeV 
for $m_{\tq}\simeq m_{\tg}$, these limits hardly affect naturalness: {\it e.g.} 
$\Delta_{\mathrm{EW}}<30$ allows for $m_{\tg}$ as high as $\sim 3-5\,$TeV and
first generation squarks are basically unconstrained so that $m_{\tq}$ values into the tens of
TeV regime are certainly allowed. 
\item {\bf Electroweakinos:} The masses of the electroweakinos -- constrained by LEP2 to have
$m_{\twpm_1}>103.5\,$GeV -- are now also just beginning to be constrained by LHC8 data 
(subject to certain model assumptions).
Some constrained scenarios include 1. models with the gaugino mass unification assumption such that sub-TeV gluinos
or first/second generation squarks would be produced strongly and then cascade-decay into electroweakinos, 
2. in conjuction with light sleptons with $m_{\tell_{L}}<m_{\tell_R}$ where $m_{\twpm_1,\tz_2}>m_{\tell}>m_{\tz_1}$ and
3. direct $\twpm_1 \tz_2$ production with decay to trileptons~\cite{Baer:1994nr} with a not-too-small 
$m_{\twpm_1,\tz_2}-m_{\tz_1}$ mass gap.
In models with light higgsinos, as motivated by electroweak naturalness, the
$m_{\tz_1}$, $m_{\tz_2}$ and $m_{\twpm_1}$ can very well be below $300\,$GeV due to their compressed
spectrum: such events are difficult to see at LHC due to prodigious QCD and EW backgrounds.
Several of the scenarios proposed below exhibit such a pattern for the light electroweakinos. 
The heavier electroweakinos, currently unconstrained by LHC8 searches, may be visible at LHC14 
in models with light higgsinos via same-sign diboson production~\cite{Baer:2013yha}: 
$pp\to\twpm_2 \tz_4\to (W^\pm\tz_2)+(W^\pm\twpm_1) +\eslt$.
The proposed benchmarks cover various options in this respect.
\item {\bf Sleptons:} The most important indication for light sleptons is still $(g-2)_\mu$. 
They are only now beginning to be constrained directly~\cite{Baer:1993ew} by LHC8 data.
If a common matter scalar mass $m_0$ at the GUT-scale is assumed, 
then the stringent LHC8 bounds on first and second generation squarks also imply rather 
heavy sleptons. 
Most of the scenarios below have heavy sleptons and thus do not explain the $(g-2)_\mu$ anomaly. 
If non-universality of matter scalars is assumed, then the slepton masses are completely unconstrained 
and all sleptons could still lie within reach of the ILC, as illustrated by the STC and NMH benchmarks described 
below: both these scenarios allow for perfect matches to the observed  $(g-2)_\mu$ value. 
In natural SUSY -- while the first two slepton generations are expected to be heavy -- 
the $\ttau_1$ can still be quite light in accord with the required light top and bottom squarks. 
\item {\bf Third generation squarks:} Direct limits on the third generation squarks from LHC8
are becoming increasingly severe and are tightly constraining natural SUSY scenarios (although not 
radiatively-driven natural SUSY). However, if the $\tst_1-\tz_1$ mass gap is small, 
as expected in stop co-annihilation scenarios, then the top squark could very well be in
the regime expected from naturalness and be accessible to ILC searches. 
The natural SUSY benchmark and the STC benchmark described in Subsections~\ref{sec:NS} 
and~\ref{sec:tdr1} give examples with light $\tst_1$ and possibly $\tb_1$ and $\tst_2$.
\item {\bf SUSY Higgses:} The SM-like properties of the newly discovered $125\,$GeV Higgs scalar
suggests that the other SUSY Higgses should be rather heavy. 
Nevertheless, we present in section~\ref{sec:nuhm2} a NUHM2 scenario with light $A$, $H$ and $H^\pm$ and in
section~\ref{sec:LMH} a case where the heavy $CP$-even Higgs boson has $m_H \simeq 125\,$GeV. Also,
the STC benchmark features heavy Higgses which should be observable at a $1\,$TeV $e^+e^-$ collider. 
\end{itemize} 

Based on these observations, we propose a set of benchmark points which can be used to 
illustrate the capabilities of the ILC with respect to supersymmetry, 
and for future optimization of both machine and detector design. 
The suggested points all lie outside the limits imposed by LHC8 searches. 
Some of these scenarios might be discoverable or excluded by upcoming LHC14 searches, while others
will be extremely difficult to detect at LHC even with $3$~ab$^{-1}$ of data at $\sqrt{s}=14\,$TeV.
The spectra for all benchmarks are available online~\cite{bib:webpage} in the SUSY Les Houches Accord format.
All figures of the spectra have been obtained with {\tt PySLHA}~\cite{Buckley:2013jua}.

\subsection{Natural SUSY (NS)}
\label{sec:NS}

Natural SUSY (NS) models are characterized by~\cite{Brust:2011tb,Papucci:2011wy,Baer:2012uy2}: 
\bi
\item a superpotential higgsino mass parameter $\mu \alt 100-300\,$GeV, 
\item a sub-TeV spectrum of third generation squarks $\tst_1$, $\tst_2$ and $\tb_1$, 
\item an intermediate scale gluino $m_{\tg}\alt 1.5-3\,$TeV with $m_A\alt |\mu|\tan\beta$ and 
\item multi-TeV first/second generation matter scalars $m_{\tq ,\tell }\simeq 10-50\,$TeV.
\ei
The last point offers at least a partial decoupling solution to the SUSY flavor and $CP$ problems.  

\begin{figure}[htb]
 \begin{center}
 \includegraphics[width=0.49\linewidth]{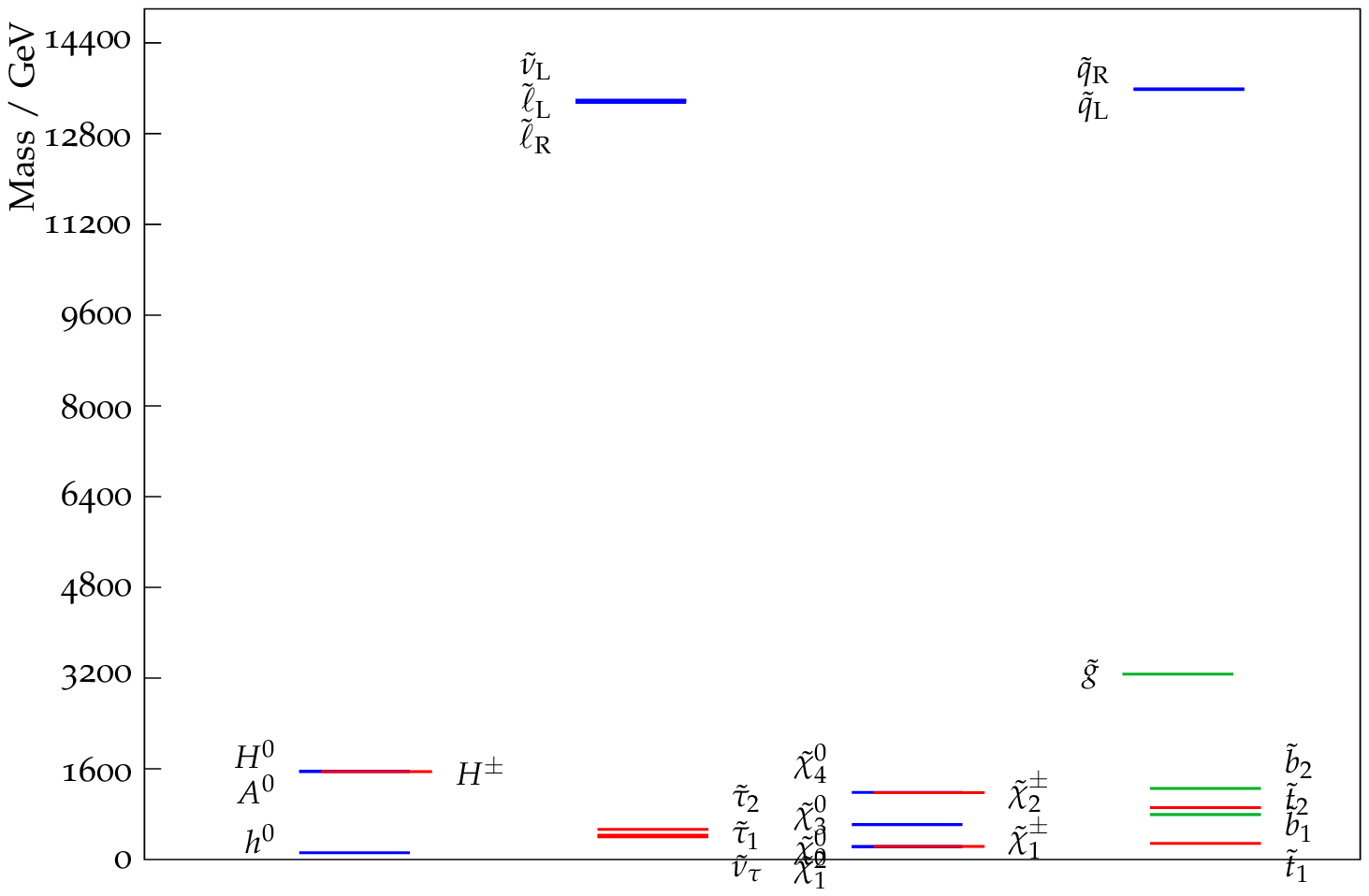}
 \hspace{0.1cm}
 \includegraphics[width=0.49\linewidth]{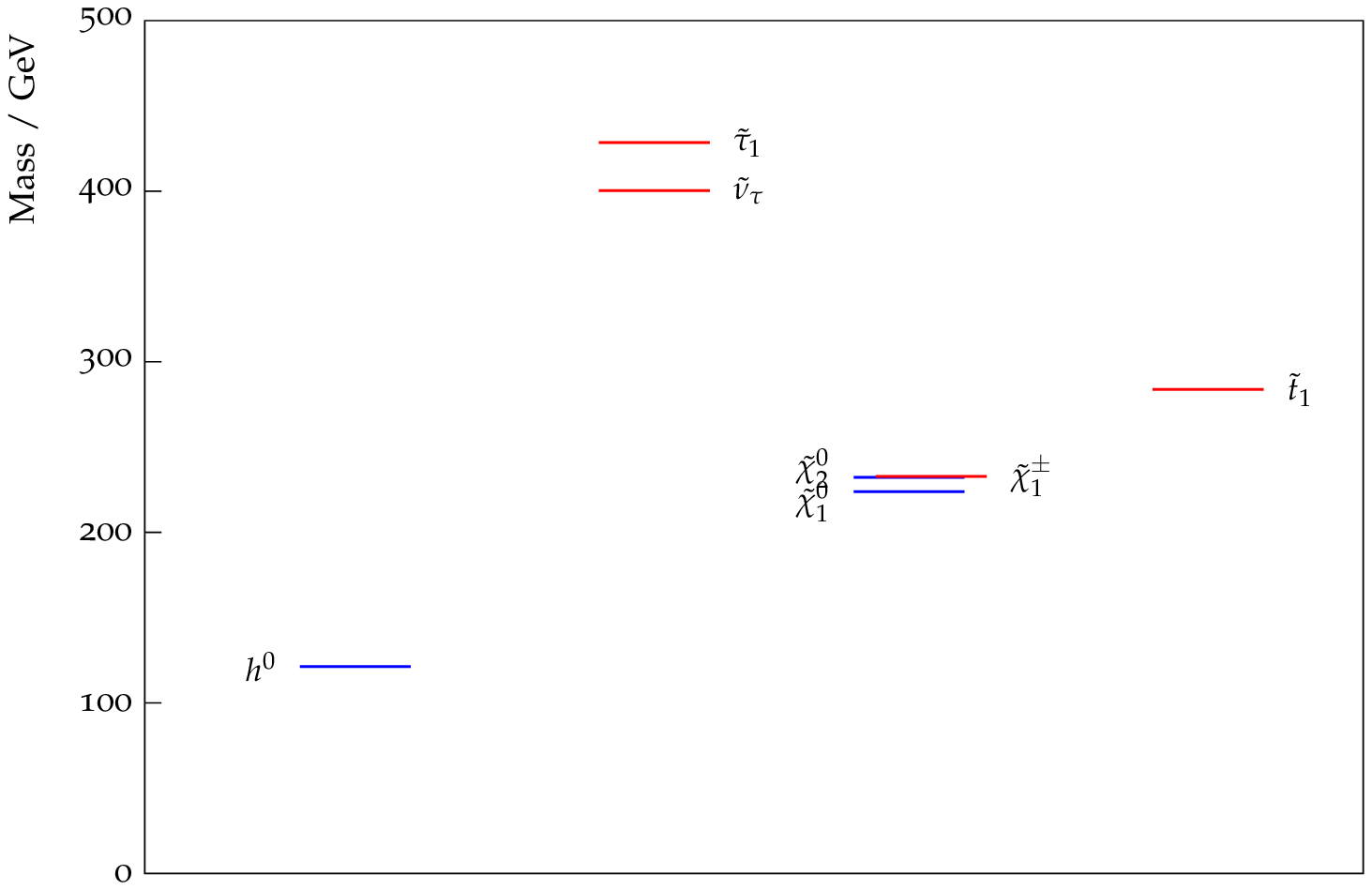}
 \end{center}
 \caption{\label{fig:massesNS} Left: Full spectrum of the natural SUSY benchmark. Right: Zoom into the spectrum below $500\,$GeV. }
\end{figure}

The suggested model parameter space which preserves gauge coupling unification is given by~\cite{Baer:2012uy2}:
\begin{equation}
m_0(1,2),\ m_0(3),\ m_{1/2},\ A_0,\ \tan\beta,\ \mu,\ m_A \;.
\label{eq:NSparams}
\end{equation}
Here, we adopt a NS benchmark point as calculated using Isasugra 7.83~\cite{Paige:2003mg} with parameters
$m_0(1,2)=13.35\,$TeV, $m_0(3)=0.76\,$TeV, $m_{1/2}=1.38\,$TeV, $A_0=-0.167\,$TeV, 
$\tan\beta =23\,$GeV, $\mu =0.225\,$TeV and $m_A=1.55\,$TeV.
The resulting mass spectrum is listed in Table~\ref{tab:bm1} and displayed in Figure~\ref{fig:massesNS} for 
all sparticles (left), and for masses below $500\,$GeV only (right).

Generic NS models with sub-TeV third generation squarks have difficulty generating 1. $m_h=125\,$GeV,
2. $BF(b\to s\gamma )$ and 3. evading LHC top/bottom squark searches.
In this case, some additional contributions to $m_h$ or a liberal allowance on theory error would be
needed for the Higgs mass along with additional flavor-violating contributions to reconcile
$BF(b\to s\gamma )$ with measurement.

Due to their small mass differences, the higgsino-like light electroweakinos will tend to look 
like missing transverse energy to the LHC. 
The next heavier particle is the $\tst_1$. 
Since the mass difference $m_{\tst_1}-m_{\tz_1}$ is less than the top mass, the decay 
$\tst_1\to b \twpm_1$ dominates, thus making the signature for $\tst_1$ pair production 
two acollinear $b$-jets plus missing transverse energy. 
The relatively compressed mass difference
$m_{\tst_1}-m_{\twpm_1}=51\,$GeV allows for evasion of LHC8 top squark search limits. 

For ILC, the spectrum of higgsino-like $\twpm_1$, $\tz_1$ and $\tz_2$ will be accessible 
for $\sqrt{s}\agt 400-600\,$GeV via $\tw_1^+\tw_1^-$ 
pair production and $\tz_1 \tz_2$ mixed production, 
albeit with a mass gap $m_{\twpm_1}-m_{\tz_1}\simeq m_{\tz_2}-m_{\tz_1}\simeq 9\,$GeV: thus, 
visible energy released from decays will be small.
Specialized cuts allowing for ILC detection of light higgsinos with small mass gaps 
have been advocated in Ref's~\cite{Baer:2003ru} and~\cite{Baer:2004zk}; there it is also demonstrated
that ILC will be able to measure the values of $\mu$ and $M_2$ and show that $|\mu |<M_2$. 

In the case of very small mass gaps, a hard ISR photon radiated from the initial state may help to lift 
the signal out of the substantial background of photon--photon induced processes. The experimental performance 
of this ISR recoil method has been evaluated recently in full simulation of the ILD detector in context of radiative
WIMP / neutralino production~\cite{Bartels:2012ex, Bartels:2012rg}.
The cross-sections are typically in the few tens of fb region~\cite{Baer:2011ec} 
and thus should be detectable in the clean ILC environment. 
Similar signatures have also been investigated in the context of AMSB for the 
TESLA TDR~\cite{AguilarSaavedra:2001rg}.

As $\sqrt{s}$ is increased past $600-800\,$GeV, then also $\tst_1\bar{\tst}_1$, $\tnu_\tau\bar{\tnu}_\tau$ 
and $\ttau_1\bar{\ttau}_1$ become successively accessible. 
This benchmark model can be converted to a model line by
varying the GUT-scale third generation mass parameter $m_0(3)$ or by varying $\mu$.
The light Higgs mass $m_h$ can be pushed as high as $\sim 124\,$GeV if larger values
of $m_0(3)$ and $|A_0|$ are selected~\cite{Baer:2012uy2}.  

\subsection{Radiatively-driven natural SUSY (RNS)}
\label{sec:hs}

\begin{figure}[htb]
 \begin{center}
 \includegraphics[width=0.49\linewidth]{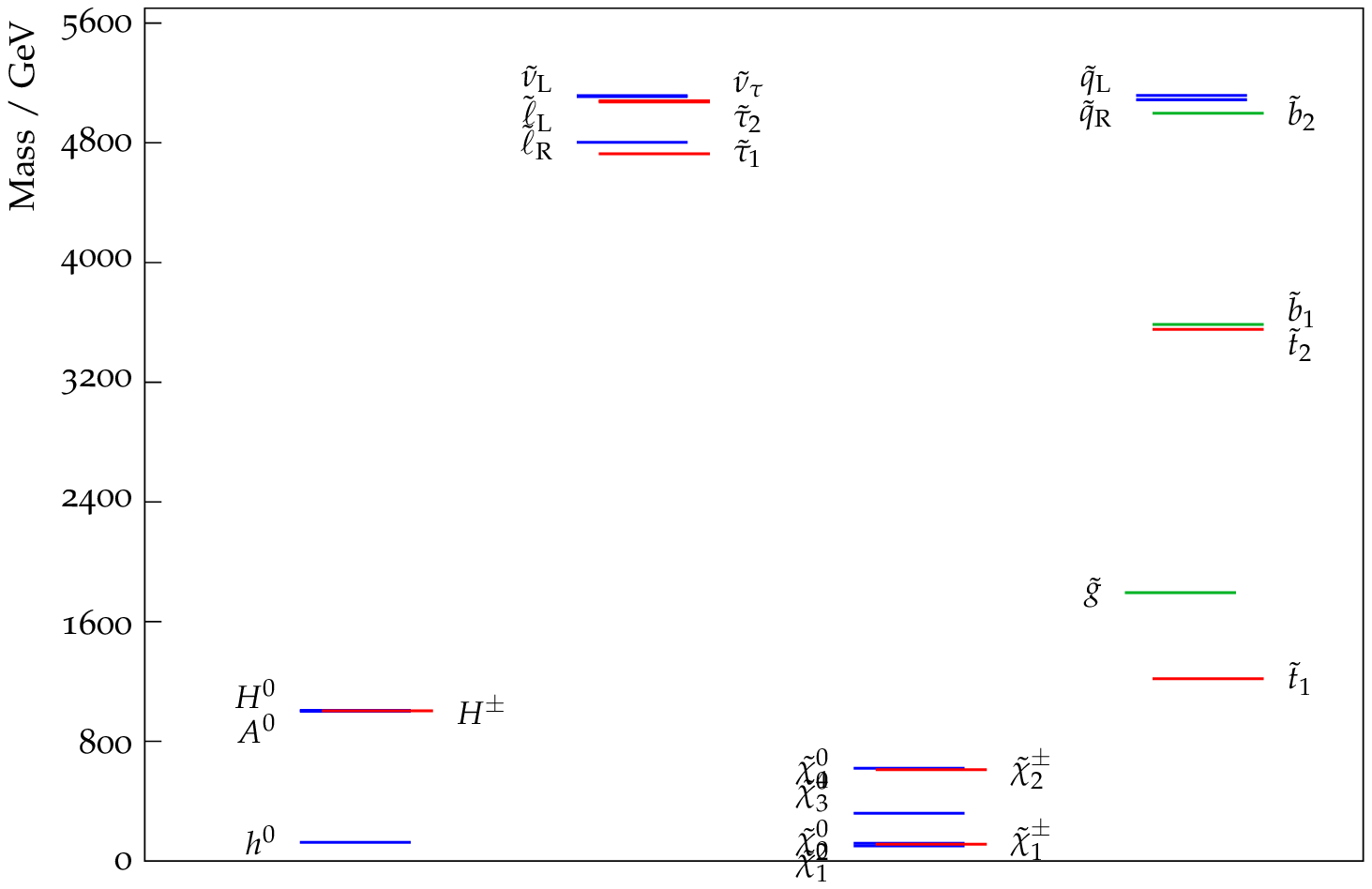}
 \hspace{0.1cm}
 \includegraphics[width=0.49\linewidth]{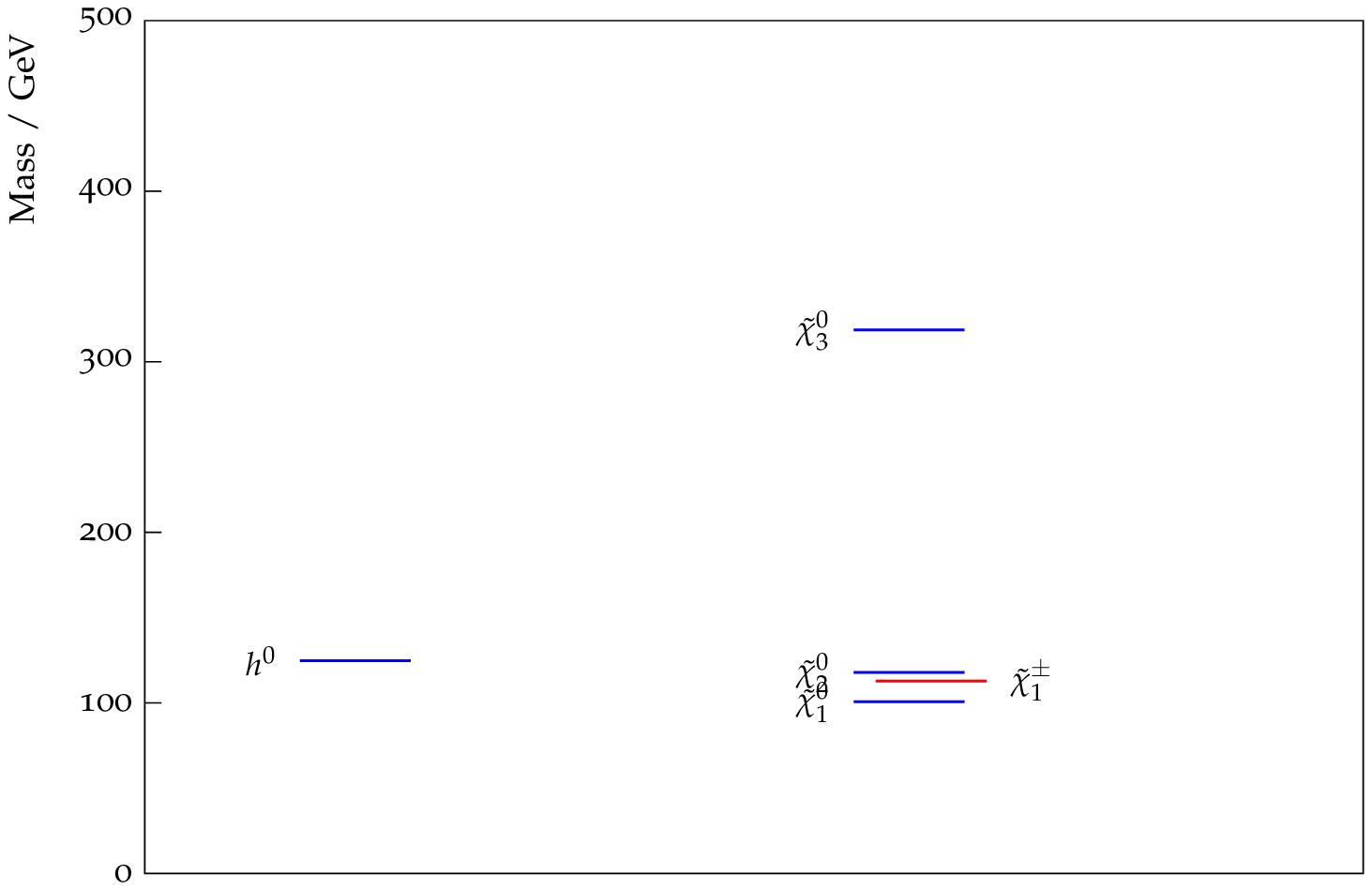}
 \end{center}
 \caption{\label{fig:massesRNS} Left: Full spectrum of the RNS benchmark. Right: Zoom into the spectrum below $500\,$GeV. }
\end{figure}

Models of RNS are motivated by trying to minimize $\Delta_{\mathrm{EW}}$ while maintaining 
gauge coupling unification and radiative EWSB due to the large top squark mass.
Low $\Delta_{\mathrm{EW}}$ is obtained by 1. requiring $\mu\simeq 100-300\,$GeV (lower is more natural),
2. $m_{H_u}^2$ should run to just small negative values at the weak scale and 3. large mixing in the top squark
sector. The large mixing suppresses the radiative corrections $\Sigma_u^u (\tst_1)$ and $\Sigma_u^u(\tst_2 )$
while increasing $ m_h$ to $125\,$GeV~\cite{Baer:2012up,Baer:2012cf}.
Thus, these models are typified by the requirement that $\mu$ alone be small~\cite{Chan:1997bi,Baer:2011ec}
since the third generation squarks can be far heavier than generic NS models with $m_{\tst_{1,2}}\simeq 1-4\,$TeV.
The RNS model can be realized within the structure of the 2-parameter non-universal
Higgs mass (NUHM2) model:
\begin{equation}
m_0,\ m_{1/2},\ A_0,\ \tan\beta,\ \mu,\ m_A \;.
\label{eq:nuhm2}
\end{equation}
Here, we adopt a benchmark point with parameters
$m_0=5\,$TeV, $m_{1/2}=0.7\,$TeV, $A_0=-8.3\,$TeV, $\tan\beta =10$
with $\mu =0.11\,$TeV and $m_A=1\,$TeV. The spectrum is given in Table~\ref{tab:bm1}.
With $\Delta_{\mathrm{EW}}=16$, we have 6.2\% EW finetuning in $m_Z$.
The resulting mass spectrum is listed in Table~\ref{tab:bm1} and displayed in Figure~\ref{fig:massesRNS} for 
all sparticles (left), and for masses below $500\,$GeV only (right).

RNS models are very difficult to detect at LHC. 
In contrast to natural SUSY, the third generation scalars are also beyond $1\,$TeV. 
While the higgsino-like light charginos and neutralinos are produced at large rates, 
the very low energy release from their decays will be hard to detect above background levels, 
making them all look like missing transverse energy. 
If $m_{\tg}\alt 1.8\,$TeV, then gluino cascade decays should be visible at LHC; however, for RNS models, 
$m_{\tg}$ can range as high as 5 TeV while maintaining $\Delta_{\mathrm{EW}}\alt 30$~\cite{Baer:2012cf}. A unique signature
for RNS at LHC14 is same-sign diboson production arising from wino pair production:
$pp\to\twpm_2\tz_4\to (W^\pm\tz_2)+(W^\pm\tw_1^\mp)$~\cite{Baer:2013yha}.

The ILC turning on at energy $\sqrt{s}\agt 250\,$GeV should already be able to detect and distinguish
$\twp_1\twm_1$ and $\tz_1\tz_2$ production as in the NS benchmark model. 
The small mass gap, angular distribution and polarization dependence of the signal cross sections 
may all be used to help establish the higgsino-like nature of the light $\twpm_1$, $\tz_2$ and $\tz_1$. 
In addition, the $\tz_3$ is accessible in mixed production with the lighter neutralinos already at 
$\sqrt{s}\agt 850\,$GeV.

Phenomenologically similar scenarios -- which are even more minimal in the sense that the $\tz_3$ 
and the $\tst_1$ are in the multi-TeV regime as well -- 
have been suggested by Br\"ummer and Buchm\"uller~\cite{Brummer:2012zc}. 
We will discuss one example in section~\ref{sec:bb}.

\subsection{NUHM2 benchmark with light $A$, $H$ and $H^\pm$}
\label{sec:nuhm2}

This benchmark point, constructed within the 2-parameter
non-universal Higgs model (NUHM2), provides a model with
relatively light $A$, $H$  and $H^\pm$ Higgs bosons while the remaining sparticles
are beyond current LHC reach. We adopt parameters
$m_0=10\,$TeV, $m_{1/2}=0.5\,$TeV, $A_0=-16\,$TeV, $\tan\beta = 7$
with $\mu =6\,$TeV and $m_A=275\,$GeV. 
With such a light $H^+$, then $tH^+$ loop contributions to $BF(b\to s\gamma )$ are large
and non-minimal flavor violation in the $b$-sector would be needed.

\begin{figure}[htb]
 \begin{center}
 \includegraphics[width=0.49\linewidth]{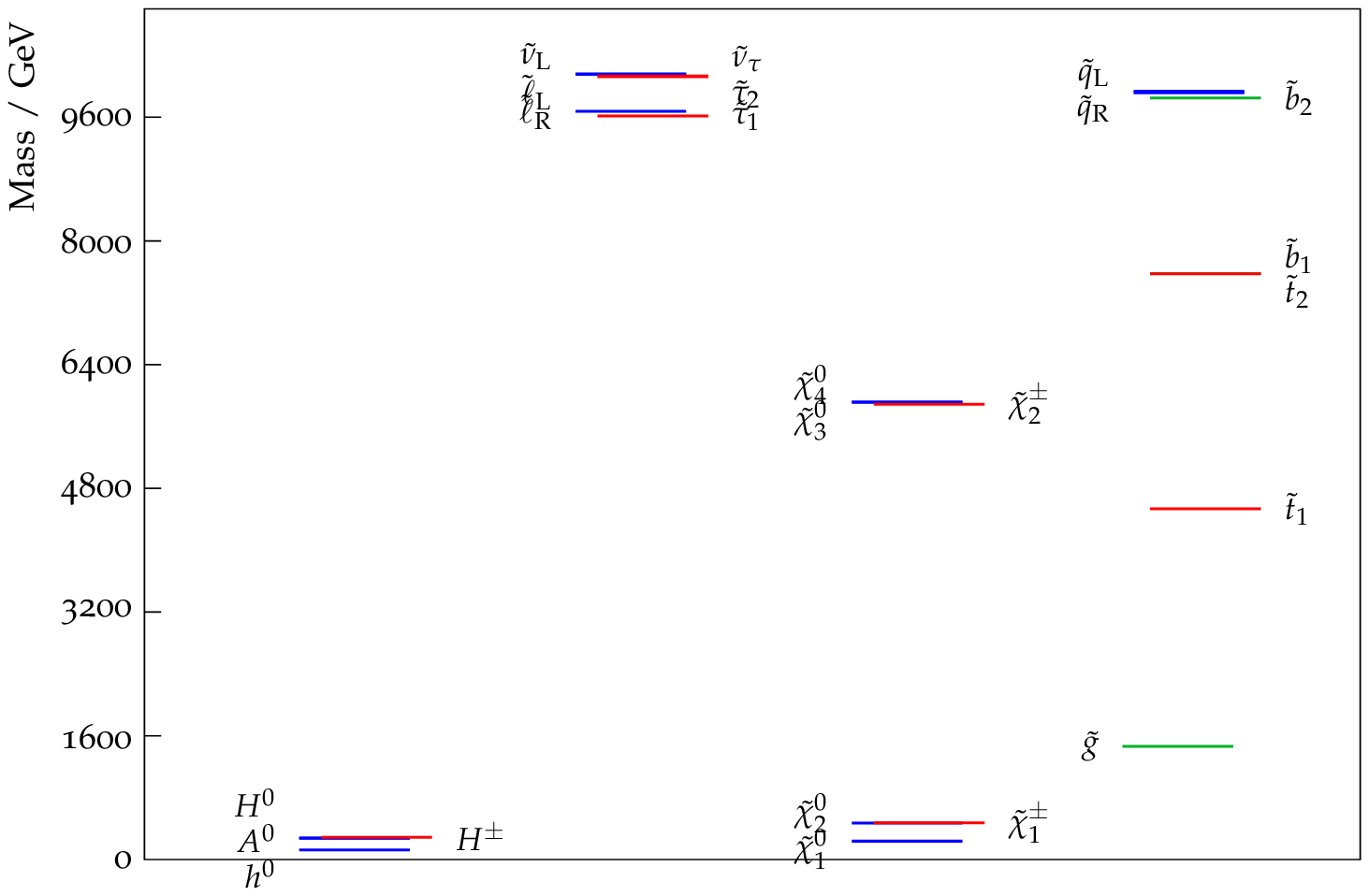}
 \hspace{0.1cm}
 \includegraphics[width=0.49\linewidth]{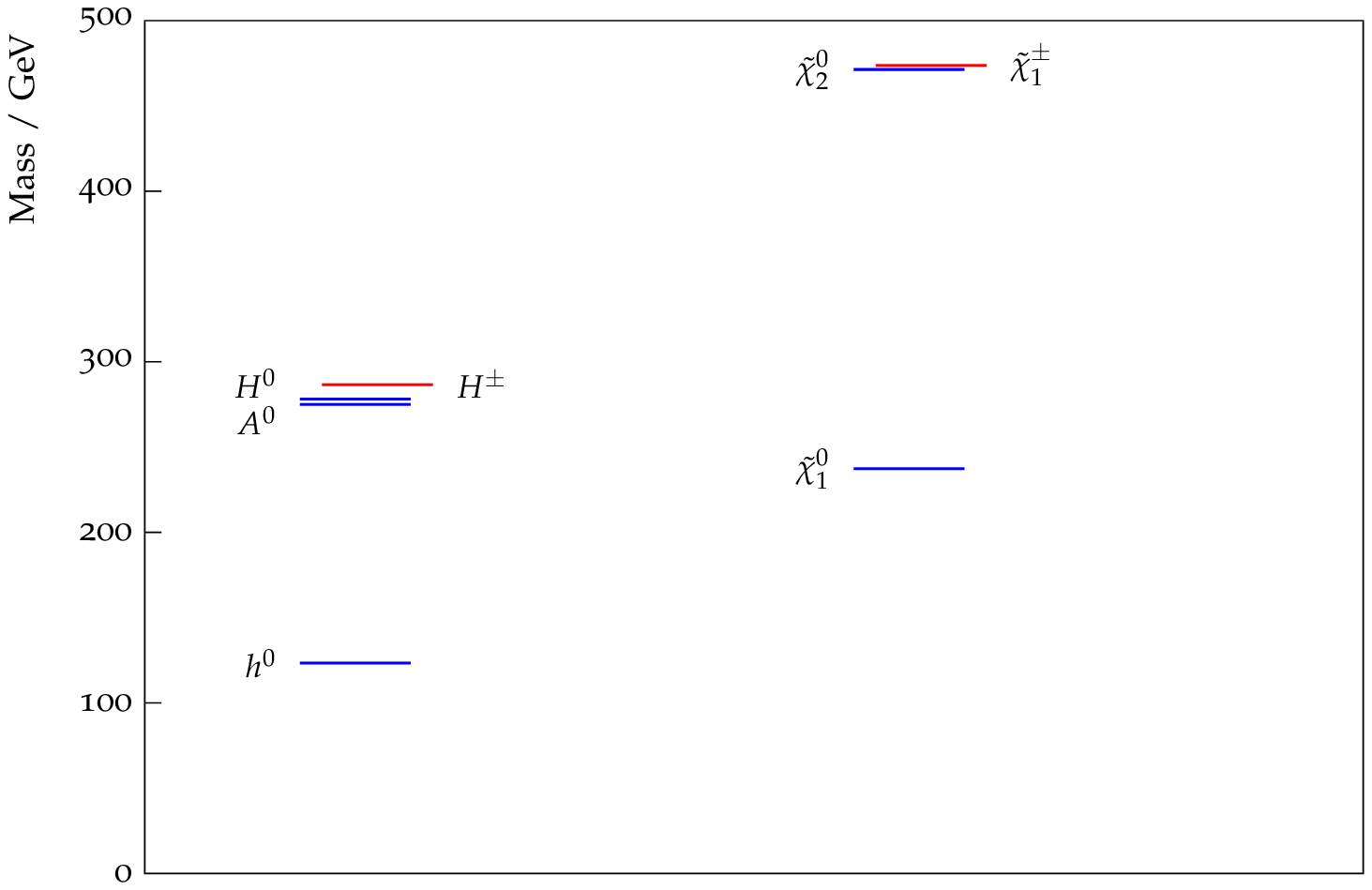}
 \end{center}
 \caption{\label{fig:massesNUHM} Left: Full spectrum of the NUHM2 benchmark. Right: Zoom into the spectrum below $500\,$GeV. }
\end{figure}

The resulting mass spectrum is listed in Table~\ref{tab:bm1} and displayed in Figure~\ref{fig:massesNUHM} for 
all sparticles (left), and for masses below $500\,$GeV only (right).

The only colored sparticles accessible to the LHC are the gluinos with $m_{\tg}=1.46\,$TeV, 
while most squarks live at around $m_{\tq}\simeq 10\,$TeV. 
The gluino decays are dominated by $\tg \ra \tz_1 t \bar{t}$ and  $\tg \ra (\twpm _1 \ra \tz_1 W^{\pm}) t b$,
and thus will require dedicated analyses for high multiplicity final states or boosted 
techniques for identifying $W$- or $t$-jets. 
The signal $pp\to\twp_1\tz_2\to Wh+\eslt\to \ell\nu_\ell+b\bar{b}+\eslt$ should ultimately be
observable at LHC14~\cite{Baer:2012ts}.
The Higgs bosons, apart from the light $CP$-even one, can most probably not be observed at the LHC 
in this low $\tan{\beta}$ and $m_A$ region~\cite{ATLASblueplot}.

At the ILC with $\sqrt{s}\simeq 0.5\,$TeV, we expect $e^+e^-\to Ah,\ ZH$ to occur at observable rates. 
As $\sqrt{s}$ rises beyond $600\,$GeV, $AH$ and $H^+H^-$ production becomes accessible
while mixed $\tz_1 \tz_2$  pair production, though accessible, is suppressed. 
At $1\,$TeV,  $\twpm_1$ and $\tz_2$ pairs will be produced in addition. 
Due to heavy sleptons and the sizable mass gap between $\twpm_1,\ \tz_2$ and the $\tz_1$, 
one expects electroweakino decays to real $W^{\pm}$ and $Z$ bosons, 
very similar to the ``Point~5'' benchmark studied in the Letter of Intents 
of the ILC experiments~\cite{Abe:2010aa, Aihara:2009ad}.

\subsection{mSUGRA/CMSSM}

Large portions of mSUGRA model parameter space are now ruled out by
direct searches for gluino and squark production at LHC8.
In addition, if one requires $m_h\simeq 124-126\,$GeV, then even larger portions of
parameter space are excluded: $m_{1/2} < 1\,$TeV (corresponding to $m_{\tg}<2.2\,$TeV)
for low $m_0$ and $m_0<2.5\,$TeV (corresponding to $m_{\tq}<2.5\,$TeV) for low $m_{1/2}$~\cite{Baer:2011ab}. 
These tight constraints rule out almost all of the co-annihilation and 
$A$-funnel annihilation regions~\cite{Baer:2011ab,Baer:2012uy}.
The HB/FP region moves to very large $m_0\agt 10\,$TeV since now $|A_0|$ must be large to accommodate 
the rather large value of $m_h$. Some remaining dark matter allowed parameter space thus remains.

\begin{figure}[htb]
 \begin{center}
 \includegraphics[width=0.49\linewidth]{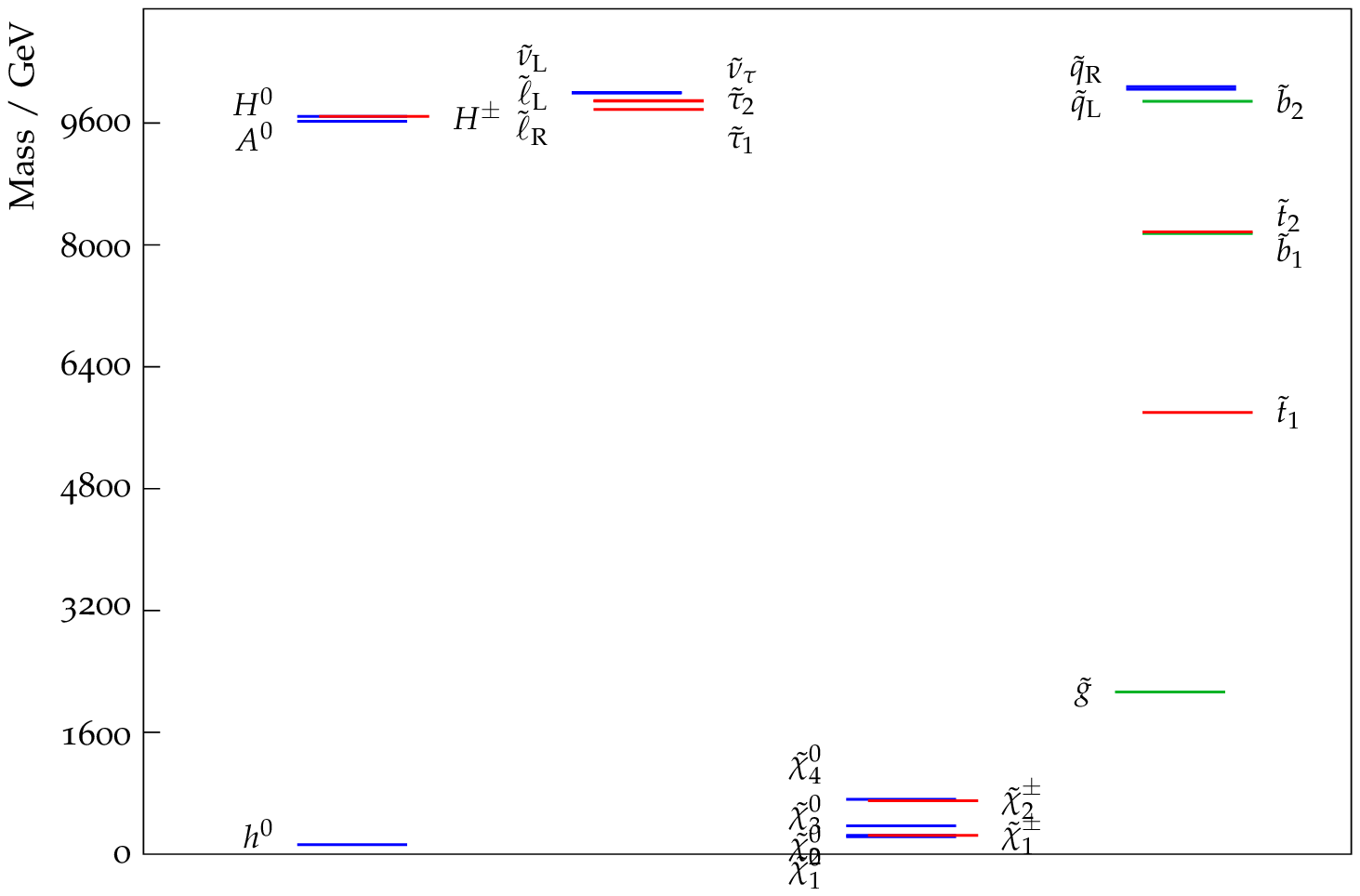}
 \hspace{0.1cm}
 \includegraphics[width=0.49\linewidth]{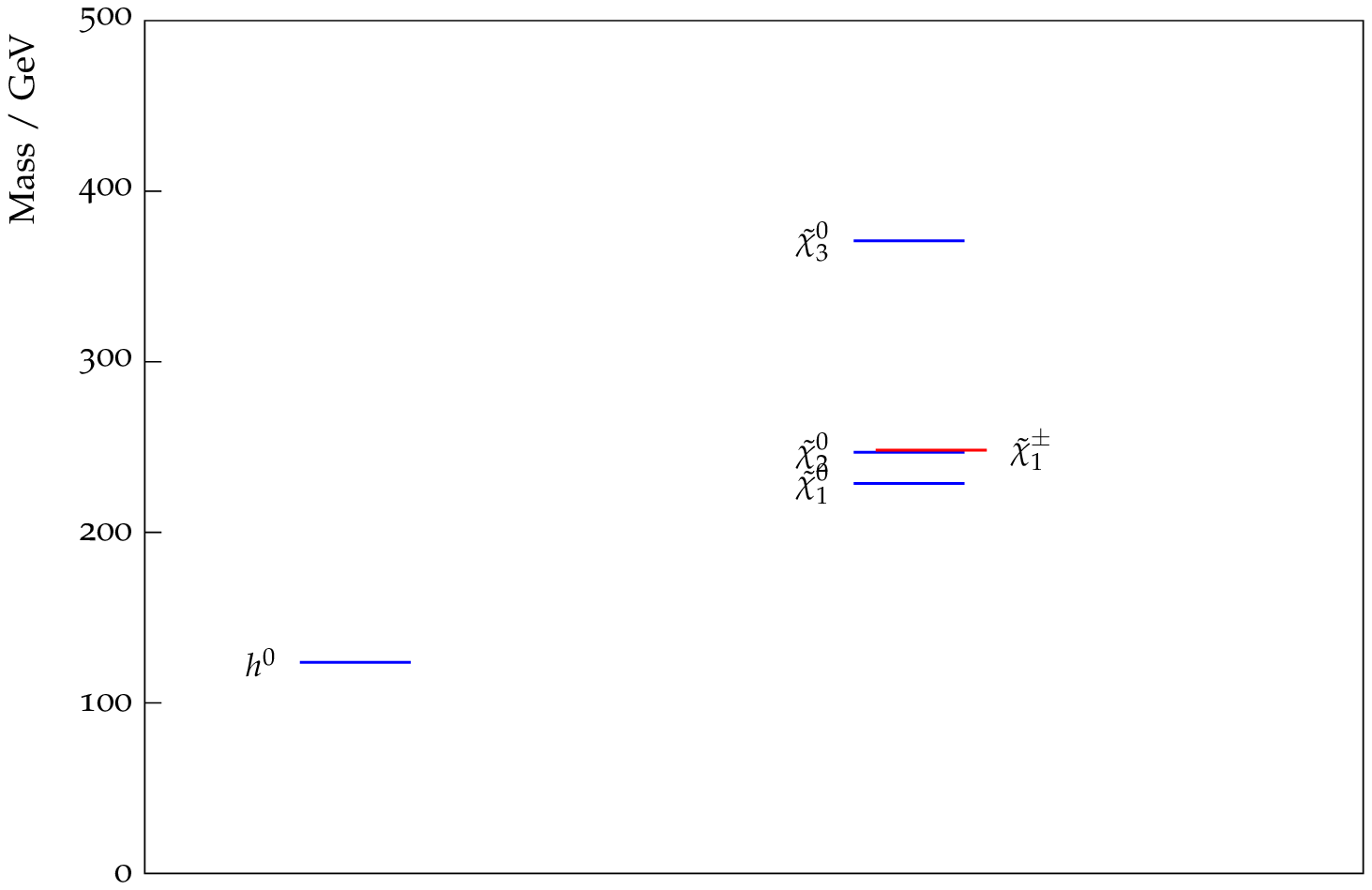}
 \end{center}
 \caption{\label{fig:massesSugra} Left: Full spectrum of the mSugra benchmark. Right: Zoom into the spectrum below $500\,$GeV. }
\end{figure}

An example is provided by an mSUGRA benchmark point with 
$m_0=10\,$TeV, $m_{1/2}= 0.8\,$TeV, $A_0=-5.45\,$TeV and $\tan\beta =15$. 
The masses are shown in Table~\ref{tab:bm1} and in Figure~\ref{fig:massesSugra} for 
all sparticles (left), and for masses below $500\,$GeV only (right).

At this point, $m_{\tg}=2130\,$GeV and $m_{\tq}\simeq 10\,$TeV so colored sparticles may be beyond LHC14 reach. 
The most promising signature for the LHC may again be $pp\to\twp_1\tz_2\to W^*h^*+\eslt$.

However, $\mu\simeq 234\,$GeV and so $m_{\twpm_1}=248\,$GeV, $m_{\tz_2}=247\,$GeV and $m_{\tz_1}=229\,$GeV. 
Thus, this point-- although still fine-tuned in the EW sector (with $\Delta_{\mathrm{EW}}=321$ 
due to $m_{\tst_{1,2}}\simeq 6-8\,$TeV)-- would allow
$\twp_1\twm_1$ and $\tz_1\tz_2$ production at ILC with $\sqrt{s}=0.5\,$TeV. 
The $\tz_1$ would be of mixed bino-higgsino variety and the $\twpm_1-\tz_1$ mass gap is just 19 GeV. 
When increasing $\sqrt{s}$ towards $1\,$TeV, the heavier neutralinos become accessible in mixed production 
and $\tz_3$ pair production.

\subsection{Non-universal gaugino masses (NUGM)}

In supergravity, gaugino masses arise from the Lagrangian term (using 4-component spinor notation)
\be
{\cal L}_F^G = -{1\over 4}e^{G/2}\frac{\partial f_{AB}^*}{\partial\hat{h}^{*j}}\left|_{\hat h\to h}\right.
\left( G^{-1}\right)_k^jG^k\bar{\lambda}_A\lambda_B
\ee
where $f_{AB}$ is the holomorphic gauge kinetic function with gauge indices $A,\ B$ 
in the adjoint representation, $\lambda_A$ are four-component gaugino fields and the $\hat{h}_m$ are hidden sector fields 
needed for breaking of supergravity. 
If $f_{AB}\sim \delta_{AB}$, then gaugino masses are expected to be universal at the high energy scale where SUSY breaking takes place. 
However, in general supergravity, $f_{AB}$ need only transform as the symmetric product of two adjoints. 
In general, gaugino masses need not be universal at any energy scale, giving rise to models with non-universal gaugino masses (NUGM).

\begin{figure}[htb]
 \begin{center}
 \includegraphics[width=0.49\linewidth]{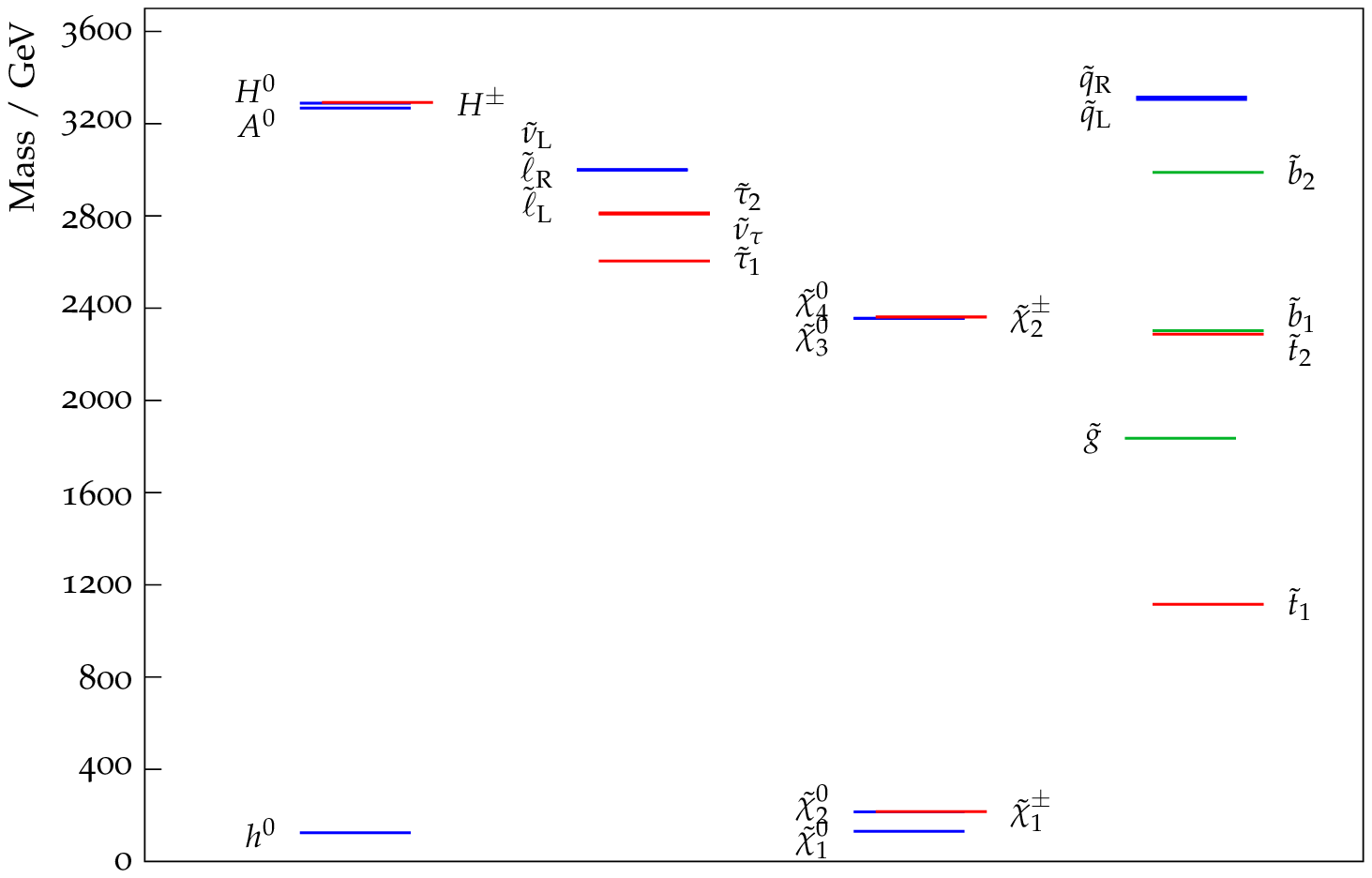}
 \hspace{0.1cm}
 \includegraphics[width=0.49\linewidth]{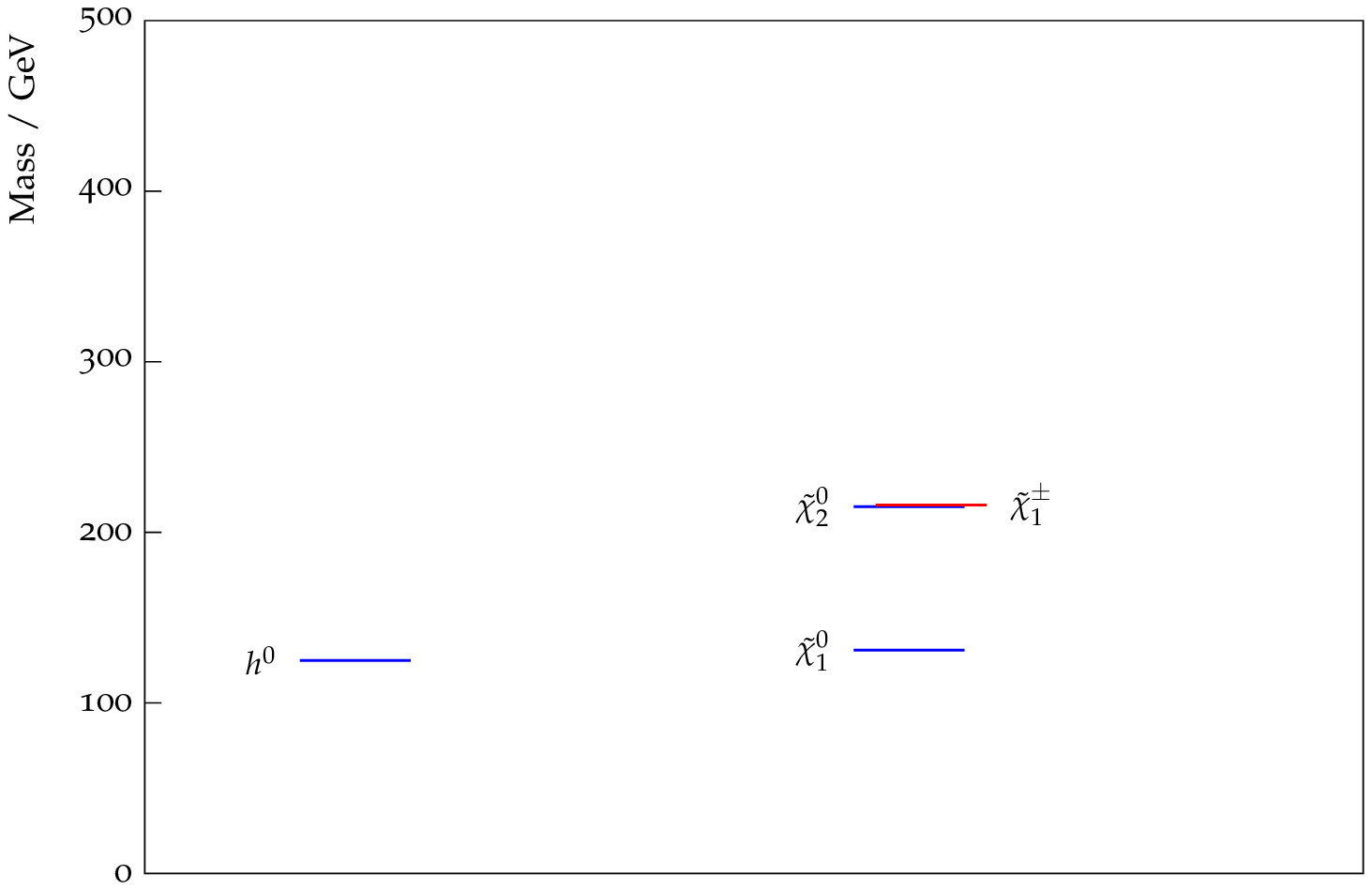}
 \end{center}
 \caption{\label{fig:massesNUGM} Left: Full spectrum of the NUGM benchmark. Right: Zoom into the spectrum below $500\,$GeV. }
\end{figure}

For a NUGM benchmark, we select a model with $m_0=3\,$TeV, $A_0=-6\,$TeV, $\tan\beta =25$ 
and $\mu >0$. We select gaugino masses at the GUT-scale as $M_1=0.3\,$TeV, $M_2=0.25\,$TeV and $M_3=0.75\,$TeV. 
The spectrum is listed in column 6 of Table~\ref{tab:bm1} and displayed in Figure~\ref{fig:massesNUGM} for 
all sparticles (left), and for masses below $500\,$GeV only (right).

With $m_{\tg}\simeq 1.8\,$TeV and $m_{\tq}\simeq 3\,$TeV, the model is clearly beyond LHC8 reach for gluinos and squarks.
The model should be testable in future LHC searches, not only in the standard jets plus missing $E_t$ analyses, 
but also via searches tailored for very high multiplicity final states and using $b$-jet tagging~\cite{Kadala:2008uy}, 
since the gluino almost exclusively decays via $\tg \ra \tst_1 t$ followed by $\tst_1 \ra \tz_1 t$. 
In addition, the production channel $pp\to\twpm_1\tz_2\to WZ+\eslt$ may be testable in the near future~\cite{Baer:2012wg}. 

The rather light spectrum of electroweak gauginos with $m_{\twpm_1}\simeq 1.6 m_{\tz_1}\simeq 216\,$GeV 
allows for chargino pair production at ILC 
followed by $\twpm_1\to\tz_1 W$ decay, yielding a $W^+W^- +\esl$ signature.
The $\tz_1\tz_2$ and $\tz_2\tz_2$ production channels tend to be suppressed, but may
offer additional search avenues albeit at low rates.

\begin{table}[h!]
\centering
\begin{tabular}{lccccc}
\hline
\hline
{\rm PMQ}                 & NS            & RNS   & NUHM2 & mSUGRA & NUGM \\
\hline
$m_{0}(1,2)$, $m_{0}(3)$  & 13.35, 0.76   & 5.0   & 10.0   & 10.0   &  3.0 \\
$m_{1/2}$ / $M_1,M_2,M_3$ & 1.38          & 0.7   & 0.5    &  0.8   &  0.3,0.25,0.75 \\
$A_{0}$                   & -0.167        & -8.3  & -16.0  & -5.450 &  -6.0  \\
\hline                                                                      
$\tan\beta$               &  23           & 10    & 7     & 15      &  25  \\
$\mu$                     & 0.225         & 0.11  & 6.0   &  0.234  &  2.36 \\
$m_{A}$                   & 1.55          & 1.0   & 0.275 &  9.62   &  3.27 \\
\hline                                                                      
$m_h$                     & 0.121         & 0.125 & 0.123 & 0.124  &  0.125 \\
$m_{H}$                   & 1.56          & 1.0 & 0.278 &  9.69    &  3.29 \\
$m_{H^{\pm}}$              & 1.55           & 1.0 & 0.286 &  9.69    &  3.29 \\
\hline
$m_{\tilde{g}}$                & 3.27        & 1.79         & 1.46         & 2.13         &  1.835  \\
$m_{\tilde{\chi}^{\pm}_{1,2}}$  &0.233, 1.18    & 0.113, 0.610 & 0.474, 5.9    & 0.248, 0.70  &  0.216, 2.36 \\
$m_{\tilde{\chi}^0_{1,2}}$     & 0.224, 0.232 & 0.101, 0.118 & 0.237, 0.471 & 0.229, 0.247   &  0.131, 0.215 \\
$m_{\tilde{\chi}^0_{3,4}}$     &0.616, 1.18   &0.319, 0.620  & 5.9, 5.9   & 0.371, 0.72      &  2.36, 2.36  \\
\hline 
$m_{ \tilde{u}_{L,R}}$         &13.58, 13.59  & 5.1, 5.3   & 9.9, 10.2   & 10.0, 10.1      &  3.30, 3.31  \\
$m_{\tilde{t}_{1,2}}$          &0.284, 0.914  & 1.22, 3.55  & 4.53, 7.57   & 5.80, 8.17    &  1.11, 2.29 \\
\hline 
$m_{ \tilde{d}_{L,R}}$         &13.6, 13.6  & 5.1, 5.1   & 9.9, 9.9   & 10.0, 10.1         &  3.30, 3.31  \\
$m_{\tilde{b}_{1,2}}$          &0.793, 1.25   & 3.6, 5.0   & 7.58, 9.85   & 8.15, 9.88    &   2.30, 2.99  \\
\hline
$m_{ \tilde{e}_{L,R}}$         &13.4, 13.3    & 5.1, 4.8   & 10.1, 9.67   & 9.99, 9.99    &  3.0, 3.0  \\
$m_{\tilde{\tau}_{1,2}}$       &0.43, 0.53    & 4.7, 5.1   & 9.61, 10.1   & 9.78, 9.89    &  2.60, 2.81  \\
\hline  
$\Omega_{\tz}^{\mathrm{std}}h^2$        &0.007         & 0.008        & 39        & 0.02         &  1085 \\
$\langle\sigma v\rangle\times 10^{25}\ [\mathrm{cm^3/s}]$ & 5.5 & 2.3 & 0.0005  & 1.5 
& 2.0$\times 10^{-7}$ \\
$\sigma^{\mathrm{SI}}(\tz p)\times 10^{9}$ [pb]        & 2.4  & 8.4  & 0.005  & 10.0               &  0.0004 \\
\hline
$a_\mu^{\mathrm{SUSY}} \times 10^{10}$                 & 0.04 & 0.07 & 0.008 & 0.04                &  0.45  \\
$BF(b\rightarrow s\gamma )\times 10^4$        & 1.8  & 3.3  & 4.6  & 3.05               &  2.95  \\
$BF(B_S\rightarrow \mu\bar{\mu})\times 10^9$   & 4.3  & 3.8  & 4.0  & 3.8                &   3.9  \\
$BF(B_u\rightarrow \tau\nu_\tau )\times 10^4$ & 1.3  & 1.3  & 1.2  & 1.3                 &   1.3  \\
\hline
$\Delta_{\mathrm{EW}}$                                 & 23   & 16    & 8782   & 321                &  1360 \\
$\Delta_{\mathrm{HS}}$                                 & 657 & 9810 & 31053    & $2.4\times 10^{4}$ &  3529 \\
\hline
\hline
\end{tabular}
\caption{Input parameters and mass spectrum and rates for post LHC8
benchmark points $1-5$. All masses and dimensionful parameters are in TeV units. 
All values have been obtained with Isasugra.
}
\label{tab:bm1}
\end{table}

\subsection{A $\ttau$-coannihilation scenario (STC)} 
\label{sec:tdr1}

In many constrained SUSY models where slepton and squark masses are
correlated at some high energy scale, relatively light sleptons with 
mass $\sim 100-200\,$GeV are forbidden. However, if we invoke the greater 
parameter freedom of the pMSSM, then spectra with light sleptons and heavy squarks
can easily be generated. In fact, these models have a motivation in that they offer
efficient dark matter annihilation mechanisms if the $\ttau_1$ is light enough, while
they at the same time naturally reconcile the measured $(g-2)_\mu$ anomaly 
(which favors light smuons) with the measured $b\to s\gamma$ branching fraction 
(which favors rather heavy third generation squarks). 

In the pMSSM~\cite{Baer:1993ae,Djouadi:2002ze}, one inputs {\it weak scale} values of the following parameters:
1. $m_{\tg},\mu ,m_A,\tan\beta$, 2. $m_Q,m_U,m_D,m_L,m_E$ for each of the three generations,
3. gaugino masses $M_1$ and $M_2$ and 4. third generation trilinears $A_t,A_b$ and $A_\tau$.
This gives a 19 dimensional parameter space if first and second generation scalar masses are taken as
degenerate, or else a 24 dimensional parameter space for independent first, second and 
third generations.\footnote{ 
Alternatively, the $SU(3)$ gaugino mass $M_3$ may be substituted for the physical
gluino mass as an input.}
As an example, we specify the ``STC'' benchmark with the following parameters, 
all given at a scale of $1\,$TeV:

\begin{itemize}
\item Higgs sector parameters:\\
 $\tan (\beta) = 10$, $\mu = 400\,$GeV, $m_A = 400\,$GeV,
\item trilinear couplings: $A_t=A_b=A_\tau = -2.1\,$TeV,
\item gaugino mass parameters:\\
 $M_3 = 2\,$TeV, $M_2 = 210\,$GeV (yields $M_1 = 100\,$GeV via GUT relation),
\item slepton mass parameters:\\
 $m_L(1,2,3) = 205\,$GeV, $m_E(1,2,3) = 117.5\,$GeV, 
\item squark mass parameters:\\
 $m_Q(1,2) = m_D(1,2) = m_U (1,2) = 2\,$TeV, 
 $m_Q(3) = 1.5\,$TeV,  $m_U(3) = 500\,$GeV, $m_D(3) = 800\,$GeV.
\end{itemize}

\begin{figure}[htb]
 \begin{center}
 \includegraphics[width=0.49\linewidth]{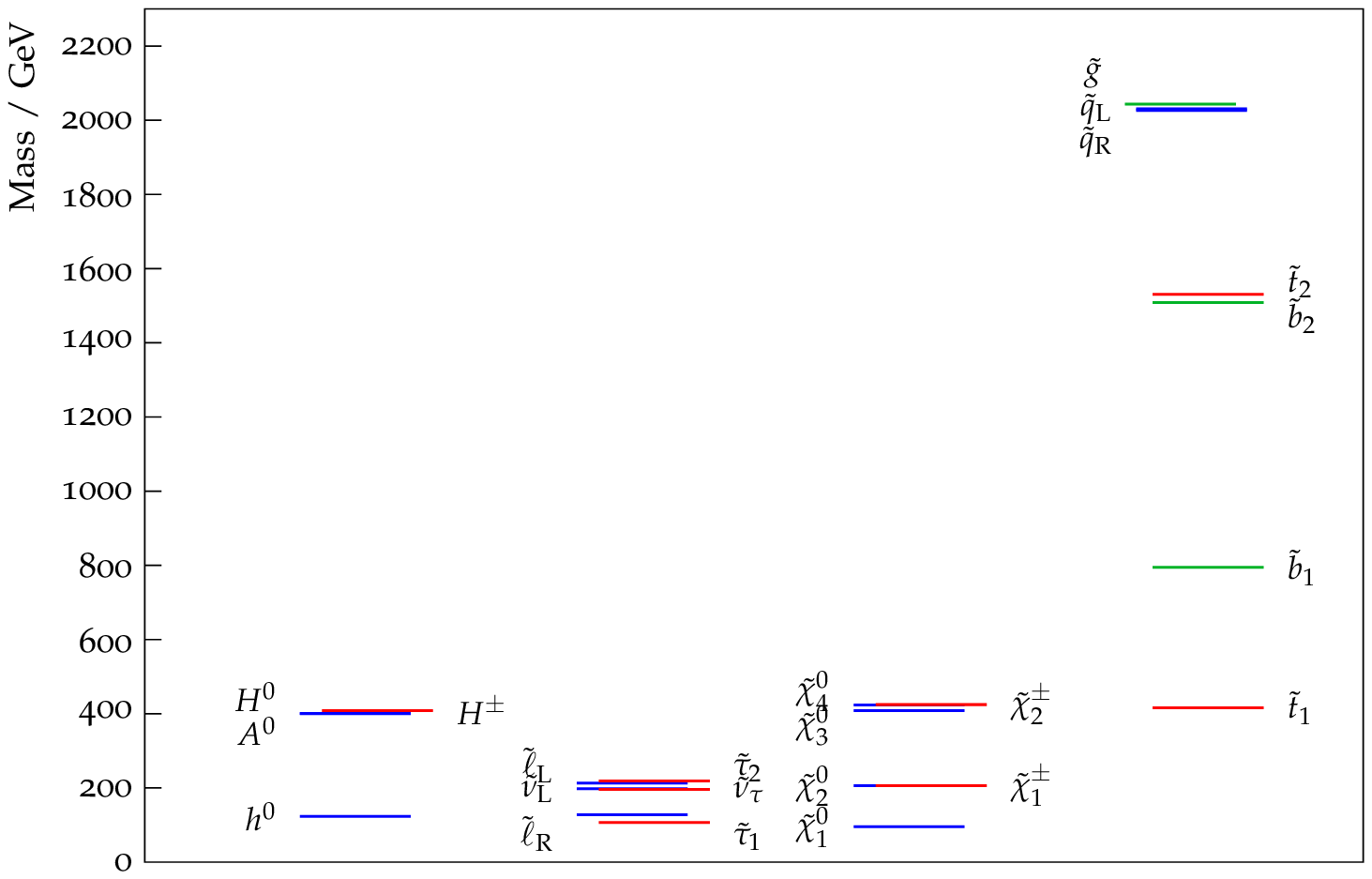}
 \hspace{0.1cm}
 \includegraphics[width=0.49\linewidth]{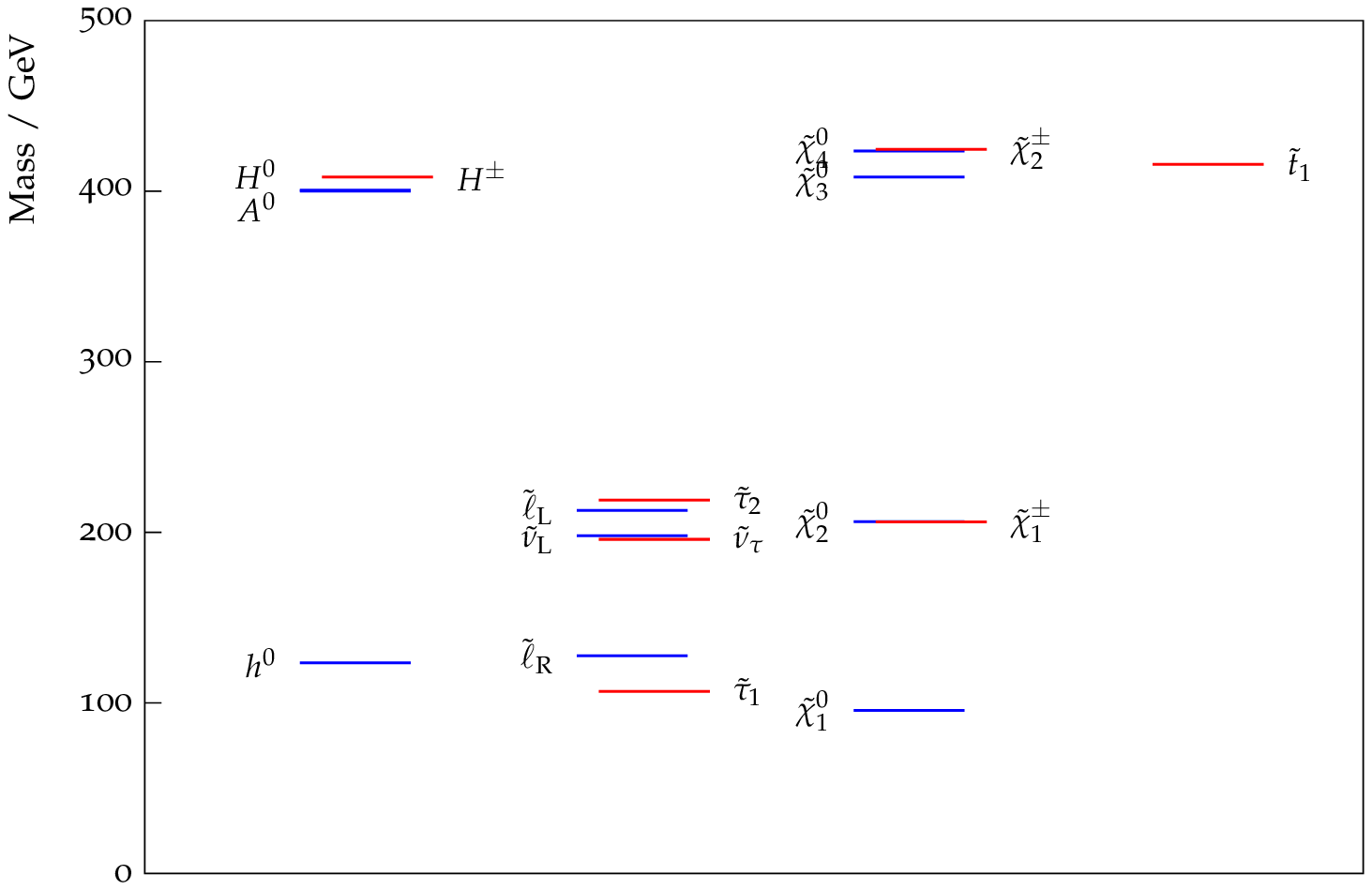}
 \end{center}
 \caption{\label{fig:massesSTC} Left: Full spectrum of the STC benchmark. Right: Zoom into the spectrum below $500\,$GeV. }
\end{figure}

Since $M_1$ and $M_2$ follow the GUT relation, there are effectively $12$ independently chosen 
parameters. The resulting sparticle masses, which have been obtained with 
SPheno~\cite{Porod:2003um, Porod:2011nf}, along with low energy and cosmic observables obtained 
from Micromegas~\cite{bib:micromegas}, are listed in 
Table~\ref{tab:bm2}  and displayed in Figure~\ref{fig:massesSTC} for 
all sparticles (left), and for masses below $500\,$GeV only (right).

The total SUSY cross-section at the LHC is about $1.5\,$pb at $8\,$TeV and doubles to 
 $3\,$pb at $14\,$TeV. The total SUSY cross-section is dominated by electroweakino production,
 which contributes about two thirds of the rate, followed by $\tst_1$ pair production. The cross-section
 for light squark production is a few fb. Thus, with masses around  $2\,$TeV, the gluino and the partners of the light quarks are beyond current LHC limits, also since the gluino decays dominantly via $\tst_1 t$ or $\tb_1 b$.  
Although light sleptons are present, the current limits on direct electroweakino 
production~\cite{bib:ATLAS_ewkino} do not cover this case due to the small mass difference between the 
$\ttau_1$ and the $\tz_1$, which leads to soft $\tau$ leptons in the chargino and 
neutralino decays instead of the searched for high $p_t$ electrons and muons.

At LHC14, a clear deviation from the SM would be observed in this scenario, but it is subject to future
study which of the numerous contributing sparticles could be identified.

At the ILC, all sleptons and the lighter electroweakinos would be observable $\sqrt{s} \alt 500\,$GeV. 
Especially the selectron pair production cross-section would be huge.
In addition, the light top squarks as well as the heavy Higgs bosons and heavy electroweakinos
would be accessible at ILC with $\sqrt{s}\simeq 1\,$TeV. It has be shown in a similar scenario that 
masses and cross-sections in the stau-sector, as well as the stau-polarisation, can be measured to a few
percent at less~\cite{Bechtle:2009em}.

Due to the large number of production processes open already at $\sqrt{s}\simeq 500\,$GeV, 
which often yield long cascades~\cite{Baer:1988kx}, STC is actually an experimentally challenging 
scenario for ILC. Therefore, it is an ideal case study to demonstrate the separation of many closely 
spaced new matter states with all the tools offered by ILC, including threshold scans and different 
beam polarization configurations, but also taking into account realistic 
assumptions on the beam energy spectrum, accelerator backgrounds and detector resolutions. 

At a center-of-mass energy of $1\,$TeV or above, the separation of the heavier electroweakinos as well 
as of the nearby $\tst_1$ and the heavy Higgs states will provide additional experimental challenges.

\subsection{Kallosh-Linde (KL), G2MSSM, spread SUSY benchmark}

While minimal anomaly-mediation seems on shaky ground due to its prediction of a
light Higgs scalar $m_h\alt 120\,$GeV, other similar models have emerged as perhaps more
compelling. One of these models -- by Kallosh and Linde 
(the KL model~\cite{Kallosh:2004yh,Linde:2011ja}) -- attempts to stabilize
stringy moduli fields via a generalization of the KKLT method~\cite{Kachru:2003aw} utilizing a racetrack
superpotential. The moduli in this theory end up superheavy and allow for 
the chaotic inflationary scenario to emerge in supergravity models. In this class of models, 
the various scalar fields have a mass of the order of the gravitino mass, with  $m_{3/2}\simeq 100\,$TeV. 
The gauginos, however, remain below the TeV scale, and adopt the usual AMSB form. 
This model is also typical of ``spread supersymmetry''~\cite{Hall:2011jd}.
Another stringy model by Acharya {\it et al.}~\cite{Acharya:2008zi} known as G2MSSM also predicts multi-TeV scalars. 
In the G2MSSM, the gauginos are again light, typically with $M_2\ll M_1\simeq M_3$ so that again a model 
with light wino-like $\twpm_1$ and $\tz_1$ emerges. 

\begin{figure}[htb]
 \begin{center}
 \includegraphics[width=0.49\linewidth]{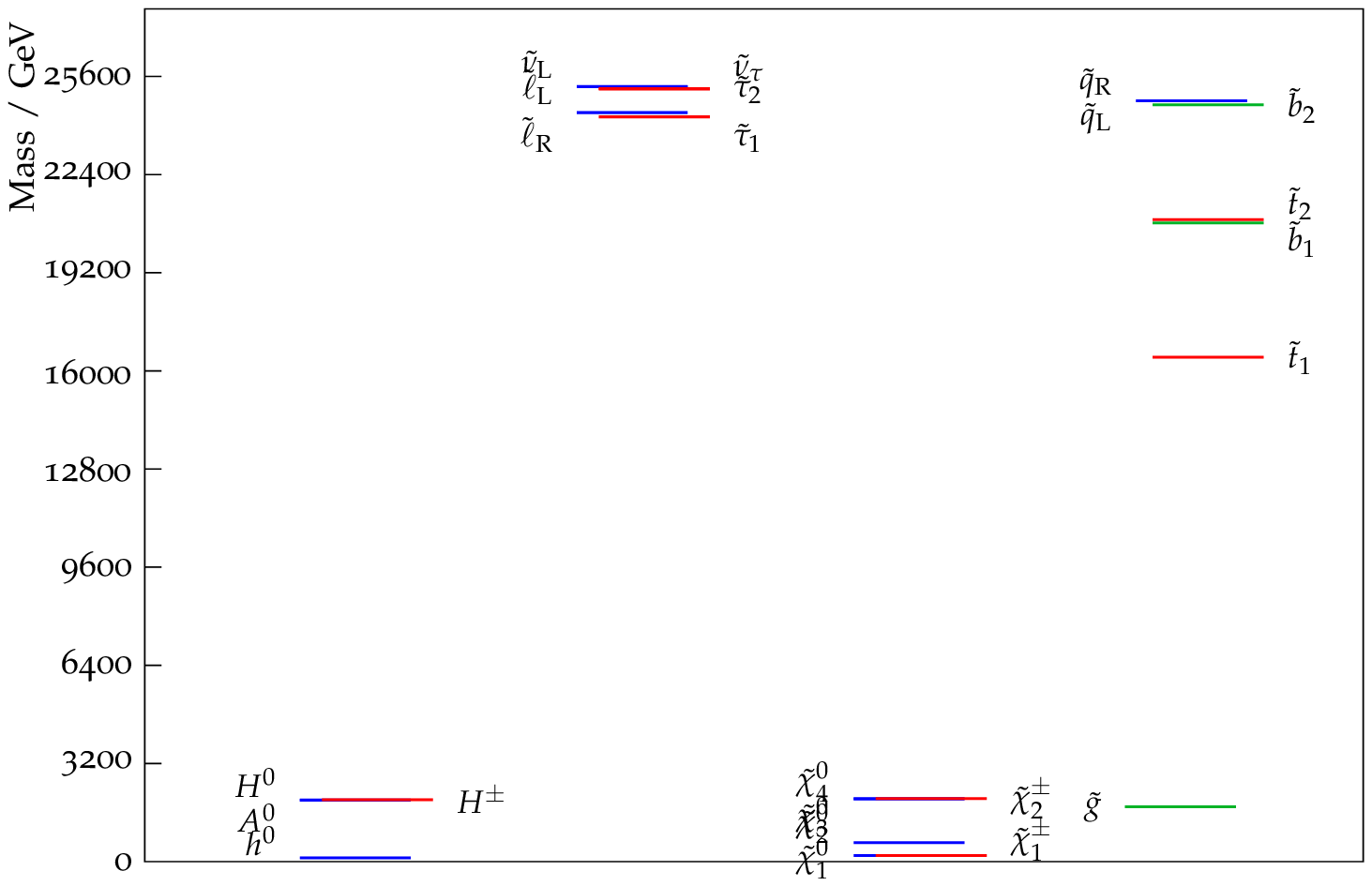}
 \hspace{0.1cm}
 \includegraphics[width=0.49\linewidth]{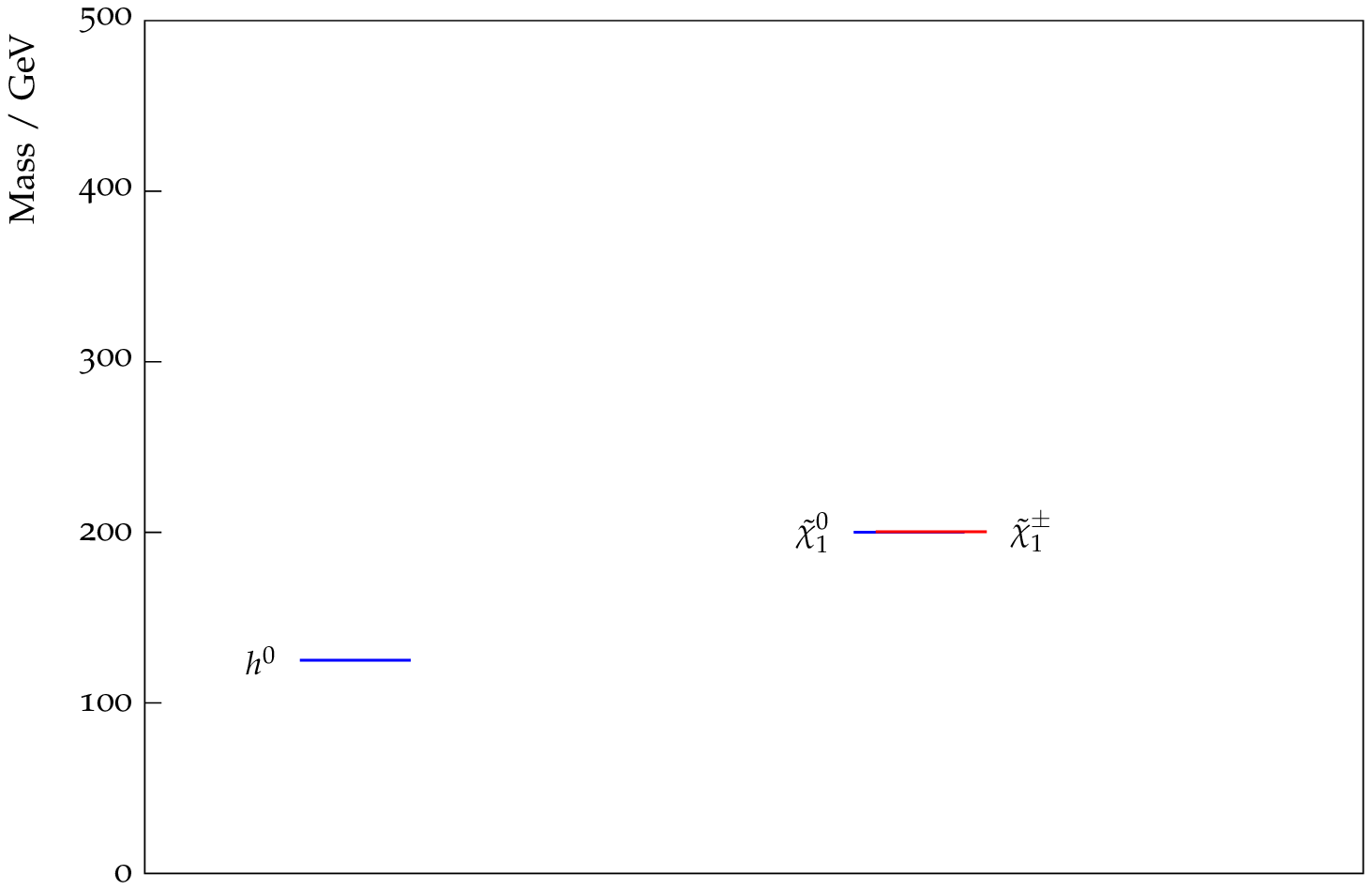}
 \end{center}
 \caption{\label{fig:massesKL} Left: Full spectrum of the KL benchmark. Right: Zoom into the spectrum below $500\,$GeV. }
\end{figure}

To model these cases, we adopt the NUHM2 model, but with non-universal gaugino masses, 
with parameters chosen as $m_0=25\,$TeV, $m_{1/2}=200\,$GeV, $A_0=0$, $\tan\beta =10$ with
$\mu =m_A=2\,$TeV. We then set GUT-scale gaugino masses to the AMSB form given by
$M_1=1320\,$GeV, $M_2=200\,$GeV and $M_3=-600\,$GeV. 

The spectrum is listed in Table~\ref{tab:bm2} and displayed in Figure~\ref{fig:massesKL} for 
all sparticles (left), and for masses below $500\,$GeV only (right). The wino-like $\tz_1$ state is
the lightest MSSM particle with mass $m_{\tz_1}=200.07\,$GeV while the wino-like lightest 
chargino has mass $m_{\twpm_1}=200.4\,$GeV. 
We also have a bino-like $\tz_2$ with $m_{\tz_2}=616.5\,$GeV and a gluino with $m_{\tg}=1788\,$GeV.
All matter scalars have mass near the 25 TeV scale, and so decouple. The light Higgs
scalar has mass $m_h=125\,$GeV.

In this case, gluino pair production may barely be accessible to LHC14 with of order
$100\,$fb$^{-1}$ of data~\cite{Baer:2003wx}. At ILC, the decay products from chargino decay will
be extremely soft. However, the wino-like chargino is then quasi-stable, flying of order centimeters before decay, leaving a highly ionizing track (HIT) which terminates upon decay into very soft decay products. 
Chargino pair production could be revealed
at ILC via initial state radiation of a hard photon, and then identification of one or more HITs, or stubs. 
In addition, if $\sqrt{s}$ is increased to $\sim 1\,$TeV, then $\tz_1\tz_2$
production opens up, although rates are expected to be small. 
In this case, one expects $\tz_2\to W\twpm_1$  or $\tz_1 h$ to occur.

\subsection{Br\"ummer-Buchm\"uller (BB) benchmark}
\label{sec:bb}

Br\"ummer and Buchm\"uller have proposed a model wherein the Fermi scale
emerges as a focus point within high scale gauge mediation~\cite{Brummer:2012zc}.
The model is inspired by GUT-scale string compactifications which frequently predict
a large number of vector-like states in incomplete GUT multiplets which may serve as
messenger fields for gauge mediated SUSY breaking which is implemented at or around the
GUT-scale. By adopting models with large numbers of messenger fields, it is found that the
weak scale emerges quite naturally from the scalar potential as a focus point from
RGE running of the soft terms. 
The soft SUSY breaking terms receive both gauge-mediated and gravity-mediated contributions.
The gauge-mediated contributions are dominant for most soft masses, while the $A$-terms and $\mu$ 
may be forbidden by symmetry.
The superpotential higgsino mass term $\mu$ emerges from gravitational interactions and
is expected to be of order the gravitino mass $\mu\simeq m_{3/2}\simeq 150-200\,$GeV.
The spectrum which emerges from the model tends to contain gluino and squark masses in the 
several TeV range so that the model is compatible with LHC constraints.
States accessible to a linear collider would include the higgsino-like light charginos $\twpm_1$
and neutralinos $\tz_{1,2}$.

For ILC studies, we adopt the benchmark model with messenger indices $(N_1,\ N_2,\ N_3)=(46,46,20)$, a 
characteristic gauge-mediated soft mass per messenger pair $m_{\mathrm{GM}}=250\,$GeV,
$\tan\beta = 48$ and weak scale values of $\mu=167\,$GeV and $m_A=4.05\,$TeV.
This results in a spectrum which is mostly on the multi-TeV scale, with only the four
higgsinos in collider range. The relevant low-scale gaugino mass parameters are $M_1 = 5.3\,$TeV and
$M_2 = 9.5\,$TeV.
\begin{figure}[htb]
 \begin{center}
 \includegraphics[width=0.49\linewidth]{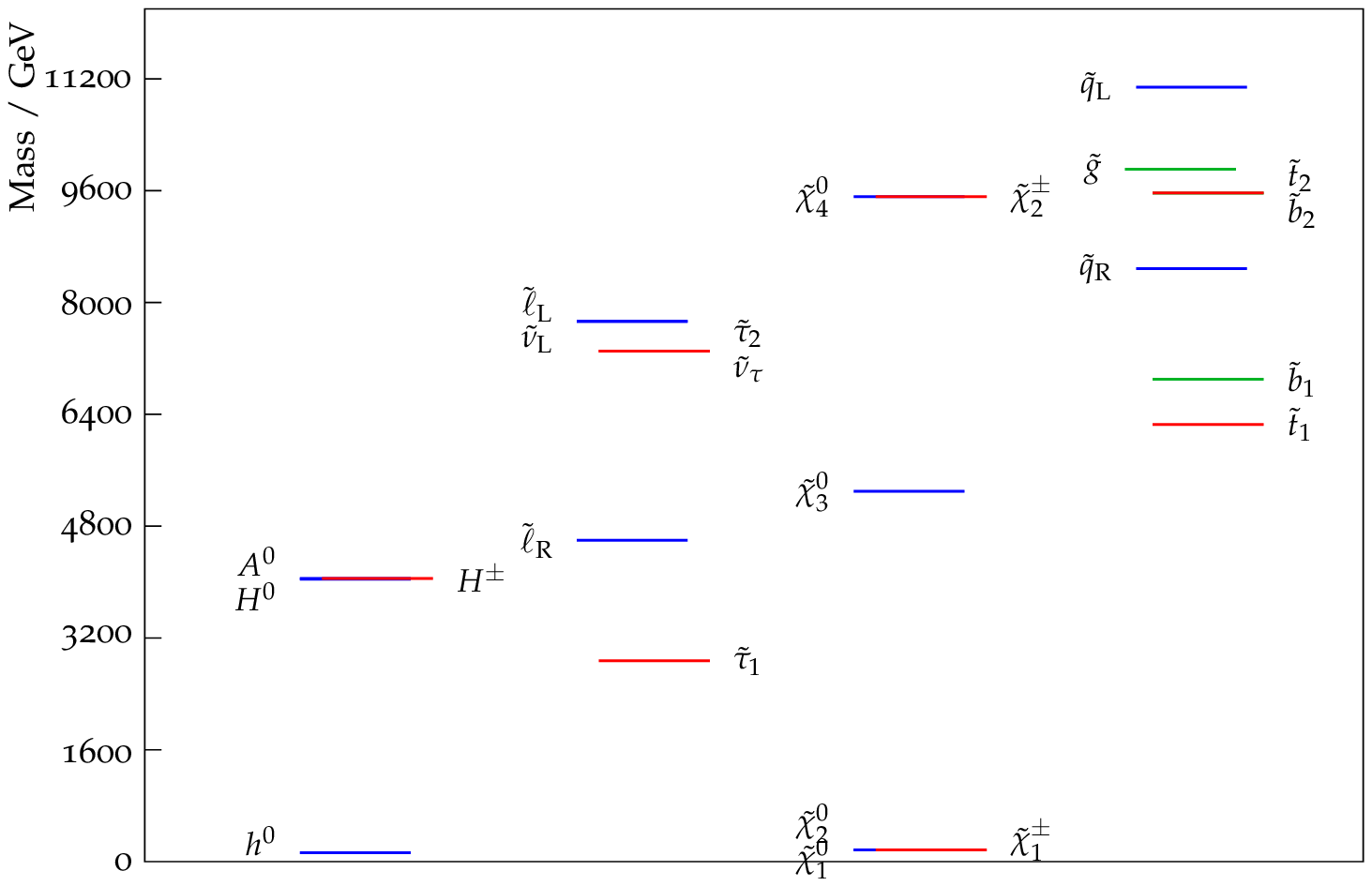}
 \hspace{0.1cm}
 \includegraphics[width=0.49\linewidth]{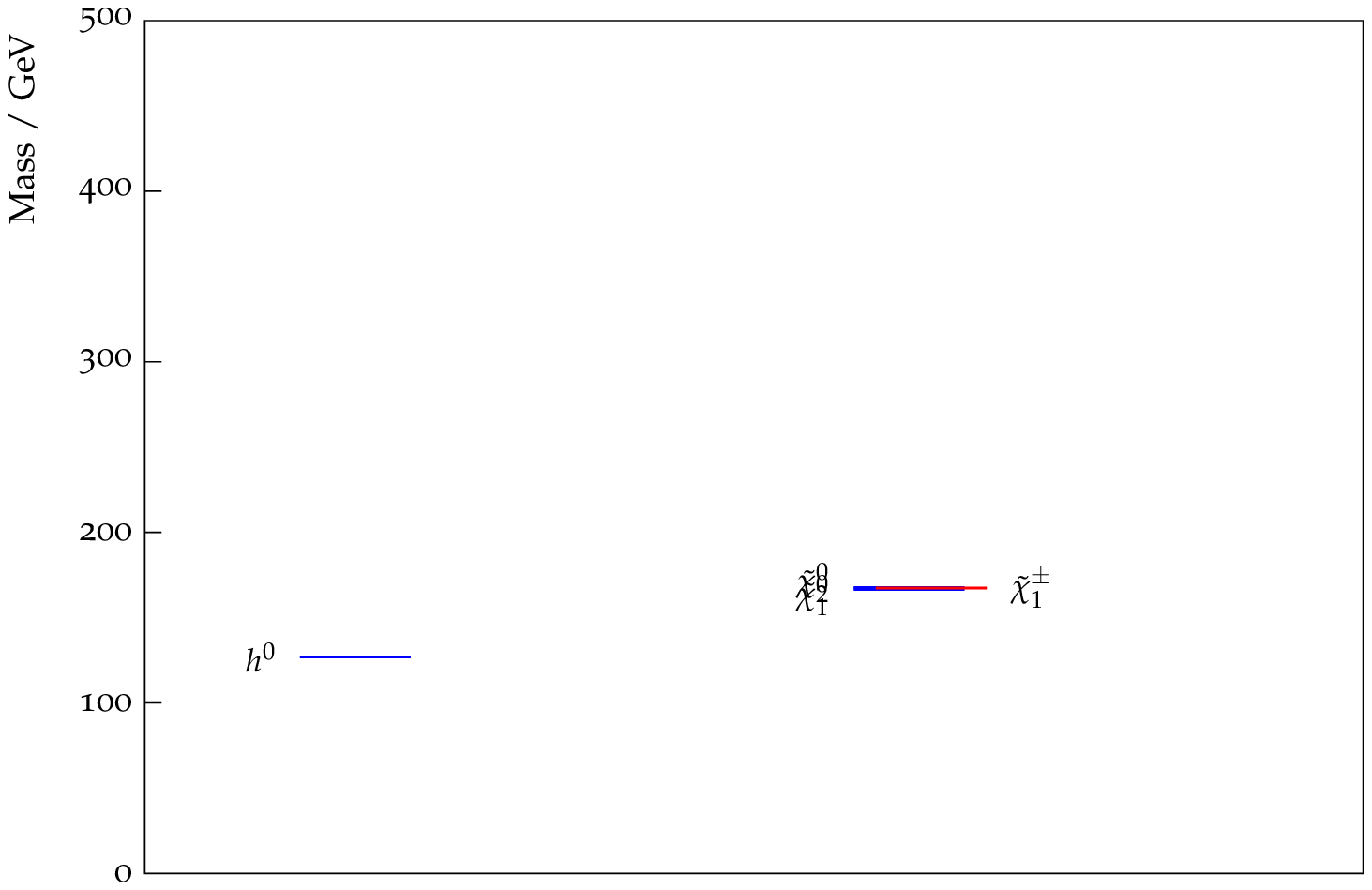}
 \end{center}
 \caption{\label{fig:massesBB} Left: Full spectrum of the BB benchmark. Right: Zoom into the spectrum below $500\,$GeV. }
\end{figure}

The spectrum,  generated from SoftSusy~\cite{bib:softsusy}, is listed in Table~\ref{tab:bm2}, 
along with low energy and cosmic observables obtained from Micromegas~\cite{bib:micromegas}. 
Figure~\ref{fig:massesKL} displays the masses for 
all sparticles (left), and for masses below $500\,$GeV only (right). 

Although the mass splitting between the Higgsinos is below $1\,$GeV, it is expected that they can
be observed at the ILC and their masses and cross-sections measured at the percent-level.

\begin{table}[h!]

\centering
\begin{tabular}{lccccc}
\hline
\hline
{\rm mass}        & STC   & KL    & BB    &  NMH & LMH \\
\hline
$\tan{\beta}$     & 10    & 10    & 48    & 20   & 6.2 \\
$\mu$             & 0.4   & 2.0   & 0.160 & 3.1  & 1.7 \\
$m_{A}$           & 0.400 & 2.0   & 4.05  & 5.36 & .110 \\
\hline
$m_h$             & 0.124 & 0.125 & 0.127 & 0.125 & .103 \\
$m_{H}$           & 0.401 & 2.013 & 4.04  & 5.34  &  .127 \\
$m_{H^{\pm}}$     & 0.408 & 2.014 & 4.05  & 5.40  &  .134\\
\hline
$m_{\tilde{g}}$                & 2.0          & 1.79          & 9.91          & 1.77  & 1.58   \\
$m_{\tilde{\chi}^{\pm}_{1,2}}$ & 0.206, 0.425 & 0.200, 2.05  & 0.167, 9.52   & 0.535, 3.1 & .211, 1.7  \\
$m_{\tilde{\chi}^0_{1,2}}$     & 0.096, 0.206 & 0.200, 0.616 & 0.167, 0.168  & 0.246, 0.533 & .095, .211\\
$m_{\tilde{\chi}^0_{3,4}}$     & 0.408, 0.424 & 2.05, 2.05    & 5.30, 9.51    & 3.06, 3.07 & 1.70, 1.70 \\
\hline 
$m_{ \tilde{u}_{L,R}}$         & 2.0, 2.03    & 24.8, 25.3    & 11.1, 8.87    & 1.62, 1.61 & 1.54, 1.54 \\
$m_{\tilde{t}_{1,2}}$          & 0.416, 1.53  & 16.4, 20.9    & 6.25, 9.57    & 2.07, 3.82 & 1.34, 1.64 \\
\hline 
$m_{ \tilde{d}_{L,R}}$         & 2.03, 2.03   & 24.8, 24.8    & 11.1, 8.49    & 1.62, 1.16 & 1.54, 1.54 \\
$m_{\tilde{b}_{1,2}}$          & 0.795, 1.51  & 20.8, 24.7    & 6.90, 9.57    & 3.84, 4.93 & 1.51, 1.54 \\
\hline
$m_{ \tilde{e}_{L,R}}$         & 0.213, 0.128 & 25.3, 24.4    & 7.73, 4.60   & 0.511, 0.252 & 1.50, 1.50 \\
$m_{\tilde{\tau}_{1,2}}$       & 0.107, 0.219 & 24.3, 25.2    & 2.87, 7.30   & 4.65, 4.85 & .126, 1.50  \\
\hline
$\Omega_{\tz}^{\mathrm{std}}h^2$                          & 0.115 & 0.0025 & 0.003 & 0.12   & 0.115        \\
$\langle\sigma v\rangle\times 10^{25}\ [cm^3/s]$ & 0.021 & 19     & 2.9   & 0.001  & 0.000012     \\
$\sigma^{\mathrm{SI}}(\tz p)\times 10^{9}$ [pb]           & 1.12  & 0.04   & 0.013 & 0.0005 & 9.3          \\
\hline
$a_\mu^{\mathrm{SUSY}} \times 10^{10}$                  & 25.6 & 0.0002 & 0.008 & 25.6 & 0.43\\
$BF(b\rightarrow s\gamma )\times 10^4$         & 3.2  & 3.2    & 3.3   & 3.2  & 5.2 \\
$BF(B_s\rightarrow \mu\bar{\mu} )\times 10^9$  & 3.4  & 3.8    & 3.1   & 3.8  & 4.4 \\
$BF(B_u\rightarrow \tau\nu_\tau )\times 10^4$  & 1.1  & 1.3    & 1.2   & 1.3  & 1.1 \\
\hline
$\Delta_{\mathrm{EW}}$ & 38 & 962                 & 275  & 2263 & 710 \\
$\Delta_{\mathrm{HS}}$ & 38 & $1.3\times 10^{5}$  & 1073 & 8295 & 710 \\
\hline
\hline
\end{tabular}
\caption{Mass spectrum and rates for post LHC8 benchmark points $6-10$. 
All masses and dimensionful parameters are in TeV units. 
All values are obtained from Isasugra for KL and NMH, while for
STC, BB and LMH  SPheno, SoftSusy and Micromegas have been used.}
\label{tab:bm2}
\end{table}

\subsection{Normal scalar mass hierarchy (NMH)}
\label{sec:nmh}

Models with a normal scalar mass hierarchy ($m_0(1)\simeq m_0(2)\ll m_0(3)$)~\cite{Baer:2004xx} are motivated 
by the attempt to reconcile the $>3\sigma$ discrepancy in $(g-2)_\mu$ (which requires
rather light sub-TeV smuons) with the lack of a large discrepancy in $BF(b\to s\gamma )$, 
which seems to require third generation squarks beyond the TeV scale. The idea here
is to require a high degree of degeneracy amongst first/second generation sfermions
in order to suppress the most stringent FCNC processes, while allowing third 
generation sfermions to be highly split, since FCNC constraints from third generation
particles are relatively mild. The {\it normal mass hierarchy} follows in that
first/second generation scalars are assumed much lighter than third generation
scalars, at least at the GUT-scale. Renormalization group running then lifts 
first/second generation squark masses to high values such that $m_{\tq}\simeq m_{\tg}$.
However, first/second generation sleptons remain in the several hundred GeV range
since they have no strong coupling.

Here, we adopt a NMH benchmark point with independent
scalar and gaugino masses at the GUT-scale.
We adopt the following parameters:
$m_5(3)\simeq m_{10}(3)=5\,$TeV, $m_{1/2}=0.63\,$TeV, $A_0=-8.5\,$TeV, $\tan\beta =20$, $\mu >0$
with $m_L(1,2)=0.21\,$TeV, $m_E(1,2)=0.387\,$TeV and $m_Q(1,2)=m_U(1,2)=m_D(1,2)\equiv
m_{10}(1,2)=0.8\,$TeV. 
We also take GUT-scale gaugino masses as $M_1=0.56\,$TeV, $M_2=0.63\,$TeV and $M_3=0.75\,$TeV.

\begin{figure}[htb]
 \begin{center}
 \includegraphics[width=0.49\linewidth]{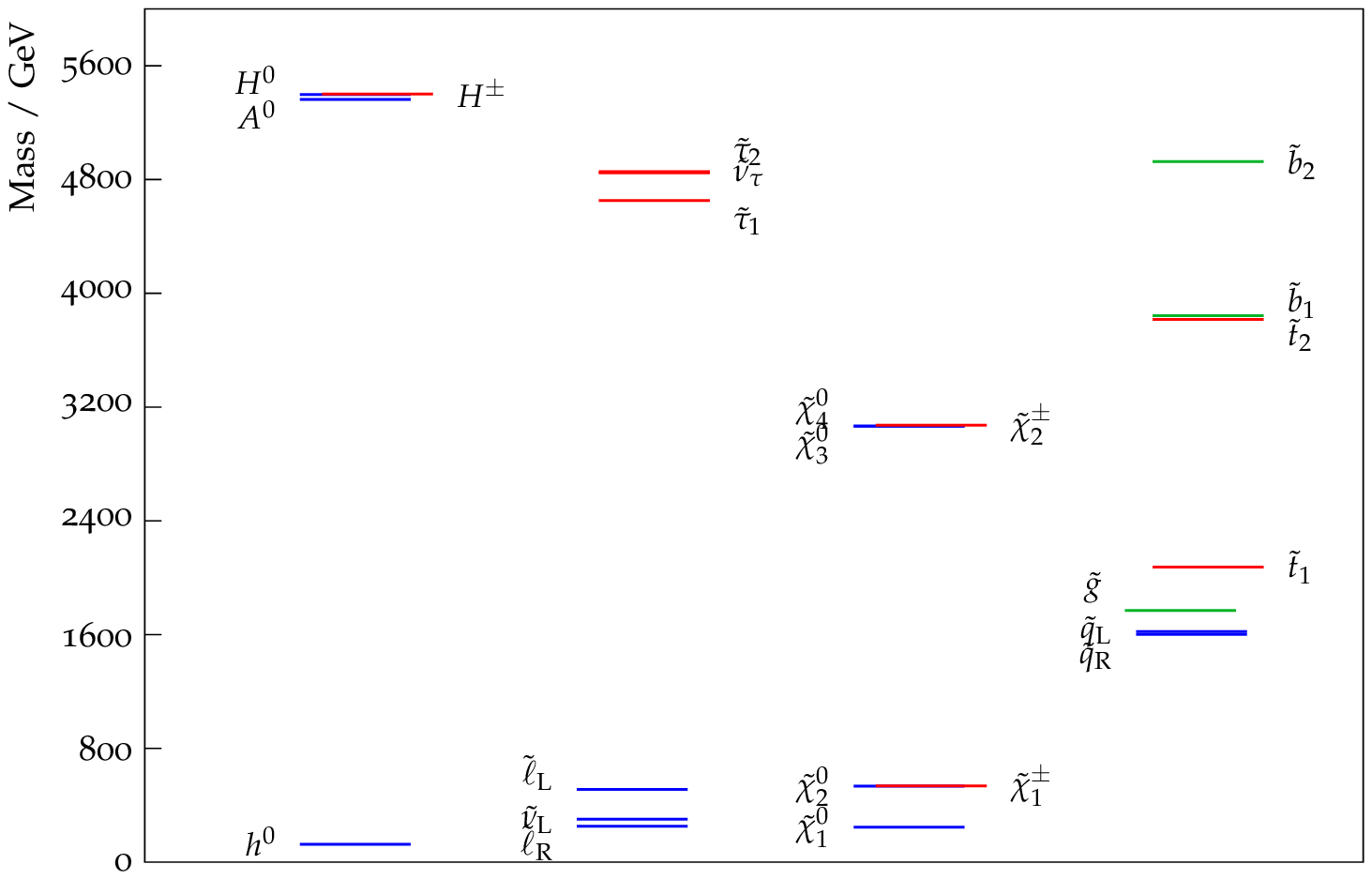}
 \hspace{0.1cm}
 \includegraphics[width=0.49\linewidth]{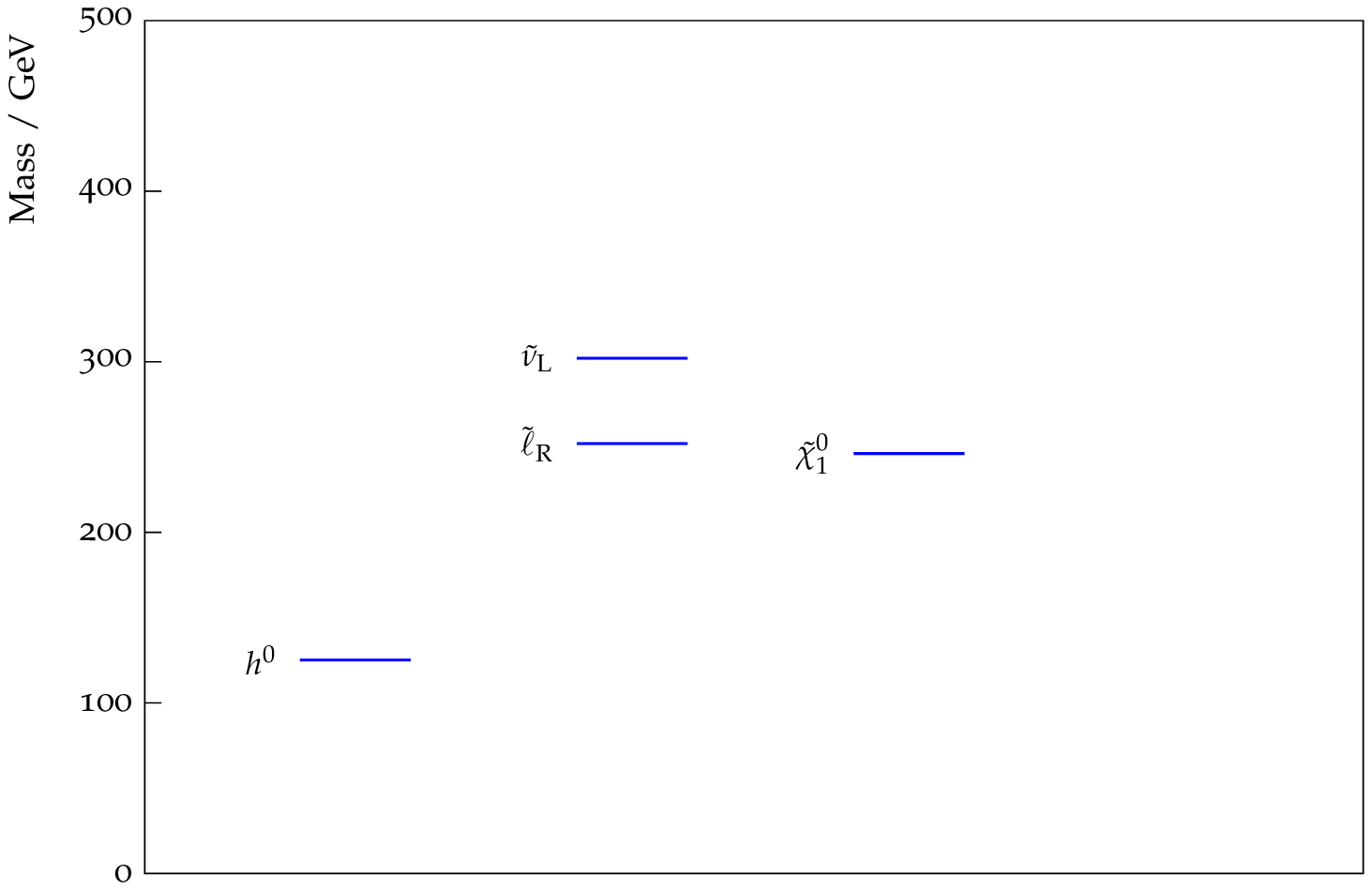}
 \end{center}
 \caption{\label{fig:massesNMH} Left: Full spectrum of the NMH benchmark. Right: Zoom into the spectrum below $500\,$GeV. }
\end{figure}

The spectrum generated using Isasugra~7.83 with non-universal
soft terms is listed in Table~\ref{tab:bm2} and displayed in Figure~\ref{fig:massesKL} for 
all sparticles (left), and for masses below $500\,$GeV only (right).

We find
$m_{\tz_1}\simeq 246\,$GeV, $m_{\te_R}\simeq m_{\tmu_R}=252\,$GeV, 
$m_{\tnu_{e,\mu L}}\simeq 302\,$GeV and $m_{\te_L}\simeq m_{\tmu_L}=512\,$GeV, as well as 
$m_h\simeq 125\,$GeV. In the colored sector, $m_{\tg} =1.77\,$TeV with $m_{\tq}\simeq 1.6\,$TeV,
so the model is compatible with LHC8 constraints, but should be testable at LHC14. 
The first and second generation squarks decay mainly into $\twpm_1 +$~jet, followed by
$\twpm_1 \ra \tnu_l l \ra \tz_1 \nu_l l$, or alternatively into $\tz_2 +$~jet, followed by
$\tz_2 \ra \tnu_l \nu_l \ra \tz_1 \nu_l \nu_l$. Thus, squark pair production will give only 2 jets,
either accompanied by just missing transverse energy or by 1 or 2 leptons. The gluino decays mostly into first or 
second generation squarks plus an additional jet. Since the $\tz_2$ decays invisibly, the only sign of direct
$\twpm_1 \tz_2$ production will be a single lepton from the the $\twpm_1$ decay plus missing transverse energy.

The model does indeed reconcile $(g-2)_\mu$ with
$BF(b\to s\gamma )$ since $\Delta a_\mu^{\mathrm{SUSY}}\simeq 26\times 10^{-10}$ and
$BF(b\to s\gamma )=3.15\times 10^{-4}$. Also, the thermal neutralino
abundance is given as $\Omega_{\tz_1}h^2\simeq 0.11$ due to neutralino-slepton
co-annihilation. An ILC with $\sqrt{s}\agt 500\,$GeV would be needed to 
access the $\te_R\bar{\te}_R$ and $\tmu_R\bar{\tmu}_R$ pair production.
These reactions would give rise to very low energy di-electron and di-muon
final states which would be challenging to extract from two-photon backgrounds.
However, since it has been demonstrated that mass differences of this size are 
manageable even in the case of $\tau$ leptons from $\ttau$ decays~\cite{Bechtle:2009em}, 
it should be feasible also in case of electrons or muons.
Since $\tnu\to\nu+\tz_1$, sneutrinos would decay invisibly, although the reaction
$e^+e^-\to\tnu_L\bar{\tnu}_L\gamma$ may be a possibility. The lack of
$\ttau^+\ttau^-$ pair production might give a hint that nature is described by 
a NMH model. 

\subsection{Low $m_H$ scenario (LMH)}
\label{sec:LMH}

For this benchmark, we assume that the $125\,$GeV particle is heavy $CP$-even Higgs boson $H$ of the MSSM. 
We adopt the recently proposed low $m_H$ benchmark scenario for the Higgs sector~\cite{Carena:2013qia}:
$m_A = 110\,$GeV, $M_{\mathrm{SUSY}} = 1.5\,$TeV, $M_2 = 200\,$GeV, 
$X_t^{\mathrm{\bar{MS}}} = 2.9 \, M_{\mathrm{SUSY}}$, $A_b=A_{\tau}=A_t$, $m_{\tg}=1.5\,$TeV, 
$M_{\tl_3} = 1\,$TeV.

\begin{figure}[htb]
 \begin{center}
 \includegraphics[width=0.49\linewidth]{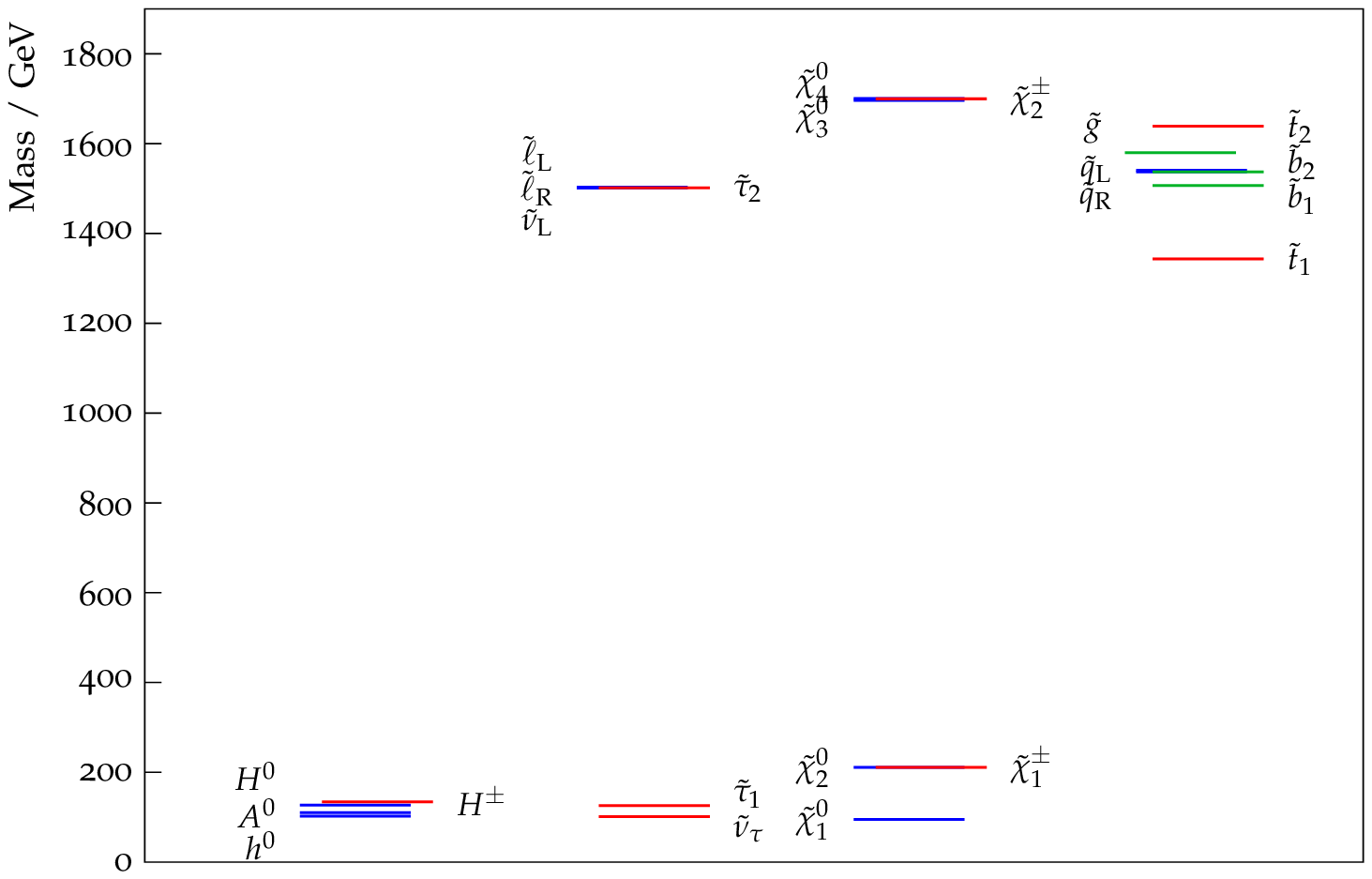}
 \hspace{0.1cm}
 \includegraphics[width=0.49\linewidth]{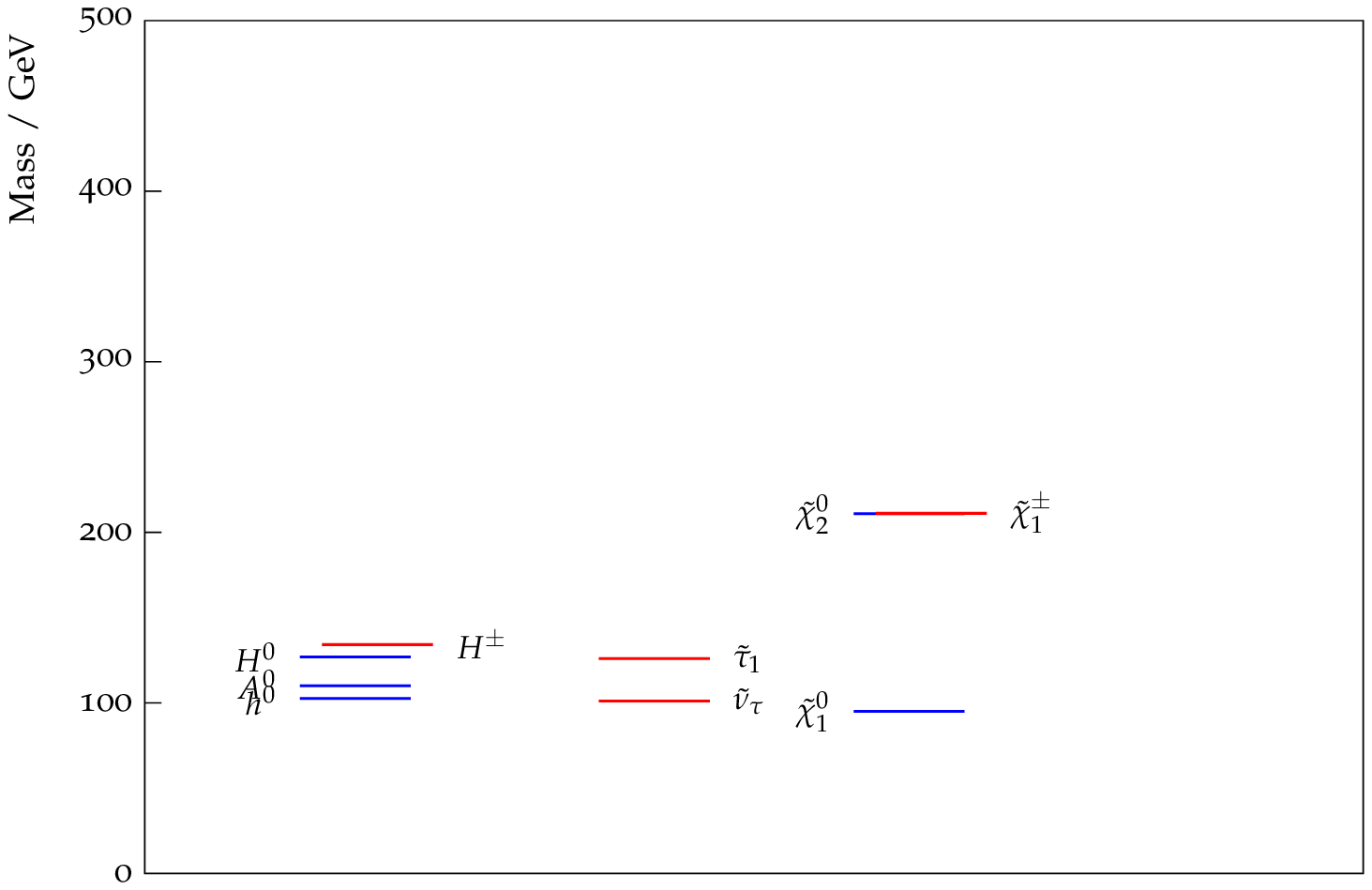}
 \end{center}
 \caption{\label{fig:massesLowMH} Left: Full spectrum of the lowMH benchmark. Right: Zoom into the spectrum below $500\,$GeV. }
\end{figure}

$\tan{\beta}$ and $\mu$ are free parameters in~\cite{Carena:2013qia}, whereas we select 
$\tan{\beta}=6.2$ and $\mu=1.7\,$TeV, which produces rates for the $H$ of at least $90\%$ 
of SM expectations. The full mass spectrum is shown in Table~\ref{tab:bm2} and illustrated 
in the left panel of Figure~\ref{fig:massesLowMH}, while the right panel is again restricted 
to sparticles with masses below $500\,$GeV.

Since the bino-like LSP annihilates only inefficiently, we lowered 
$M_{\tl_3}$ from $1\,$TeV to $113\,$GeV. As a result, the $\tnu_{\tau}$ becomes the NLSP with 
$m_{\tnu_{\tau}} = 101\,$GeV followed by the $\ttau_1$ with $m_{\ttau_1} = 126\,$GeV. Both of 
them contribute to a sufficiently high co-annihilation fraction and lower the relic density 
to the observed value. The $\ttau_1$ decays in $100\%$ of the cases into $\tau \tz_1$.

 In this channel, it has been shown that the stau mass and pair-production cross-section can be measured very precisely,
even in the more challenging situation of a smaller mass difference~\cite{Bechtle:2009em}.
The $\tnu_{\tau}$ however decays invisibly to $\nu_{\tau} \tz_1$. Since the cross-section
is too low for detecting an excess in the mono-photon signature over the SM neutrino and SUSY
$\tz_1$ pair production, it would be interesting to study if the precision on $m_{\tnu_{\tau}}$
expected from cascade decays, especially from $\twpm_1 \ra \tnu_{\tau} \tau$ and 
$\tz_2 \ra \tnu_{\tau} \nu_{\tau}$, allows to predict the relic density with precisions 
comparable to the cosmological measurements.

The light $CP$-even Higgs has a mass of $103\,$GeV; this is compatible with LEP bounds since 
it is non-SM-like but has reduced couplings to the $Z$ boson and thus the production cross-section
is smaller by about a factor $10$. The $CP$-odd Higgs and the charged Higgses have masses of 
$m_A=110\,$GeV and $m_{H^{\pm}}=134\,$GeV. Thus all Higgs bosons could be precisely studied at 
the ILC, while due to the low value of $\tan{\beta}$, they're difficult to observe at the LHC 
despite their low masses. 

\section{Conclusions}
\label{sec:conclude}

At first sight, it may appear very disconcerting that after one full
year of data taking at LHC8, with $\sim 20\,$fb$^{-1}$ per experiment, no sign of supersymmetry
is yet in sight. On the other hand, the discovery of a light higgs scalar with mass $m_h\simeq 125\,$GeV
lends indirect support to SUSY: while $m_h$ can theoretically inhabit a rather large range of values of up to $800\,$GeV 
in the Standard Model, 
the simplest supersymmetric extensions of the SM require it to lie below $\sim 135\,$GeV.
A light SUSY Higgs of mass $\sim 125$ GeV seems to require top squark masses $m_{\tst_i}\agt 1\,$TeV
along with large mixing: 
thus, the emerging overall view of the SUSY landscape seems more consistent with a super-TeV sparticle mass
spectrum than with a sub-TeV spectrum, at least as far as gluinos and squarks are concerned.
In addition, the 125 GeV Higgs signal puts a high degree of stress, as measured by SUSY naturalness, 
on many popular constrained models such as CMSSM, mAMSB and mGMSB. In the case of the MSSM, 
the Higgs signal may favor gravity-mediated SUSY breaking models since these naturally accommodate 
large mixing in the top squark sector. 

While some pre-LHC analyses (based on global fits of SUSY to a variety of data which may have been overly skewed by the
$(g-2)_\mu$ anomaly) had predicted a very light sparticle mass spectrum, 
 the presence of a multi-TeV spectrum of at least first/second generation
matter scalars was not unanticipated by many theorists. The basis of this latter statement
rests on the fact that a decoupling of first/second generation matter scalars either solves or
at least greatly ameliorates: the SUSY flavor problem, the SUSY $CP$ problem, the SUSY GUT proton decay
problem and-- in the context of gravity mediation where the gravitino mass sets the scale for
the most massive SUSY particles-- the gravitino problem. 

Such a decoupling spectrum need not be inconsistent with electroweak fine-tuning arguments.
Minimization of large log contributions to  $m_h$ implies a SUSY spectrum
including three third generation squarks with mass less than about 500 GeV: 
these are the so-called natural SUSY models. While intriguing, such a spectrum seems phenomenologically disfavored 
by the $BF(b\to s\gamma )$, by the rather high value of $m_h$, and by lack of top/bottom squark signals at LHC8.
On the theory side, minimization of large logs may be too harsh a finetuning restriction since it neglects
possible correlations amongst high scale parameters (as in focus point SUSY) which lead to natural cancellations leading to
low $m_{H_u}^2$ at the weak scale. The more conservative electroweak finetuning measure implies models with:
low $|\mu |\alt 300\,$GeV, third generation squarks with $m_{\tst_{1,2},\tb_1}\sim 1-4\,$TeV and $m_{\tg}\sim 1-5 \,$TeV.
Since first/second generation matter scalars don't enter the electroweak
scalar potential, these sparticles can indeed exist in the 10-50 TeV regime -- as required by decoupling --
without affecting fine-tuning. 
These radiatively-driven NS models allow for heavier third generation squarks while explaining how it is that the
$W,\ Z$ and $h$ masses all lie near the 100 GeV scale. 
We have presented here four different models with the required small superpotential higgsino mass $\mu$: 
NS, RNS, focus point SUSY and a non-minimal GMSB model suggested by Br\"ummer and Buchm\"uller (BB).
The spectra from these small $\mu$ models can be difficult to detect at LHC 
since the characteristic light higgsinos have a highly compressed mass spectrum.
In each of these cases, however, an ILC would easily discover the predicted light higgsino states.
In such cases, the ILC would be a {\it higgsino factory}, in addition to a Higgs factory! 

We also presented several benchmark models consistent with LHC and other constraints which predict some varied 
phenomenology. 
The NUHM2 point contains heavy matter scalars but with $A$ and $H$ Higgs bosons which would also be accessible to ILC. 
The non-universal gaugino mass (NUGM) model allows for
chargino pair production at ILC followed by $\twpm_1\to W\tz_1$ decay, leading to $W^+W^- +\esl$ events.
We also presented one benchmark point consistent with Kallosh-Linde/spread SUSY/G2MSSM models. 
In this case, matter scalars have masses
$m_{\tq,\tell}\simeq m_{3/2}\sim 25\,$TeV, but gaugino masses follow the AMSB pattern, with the $\twpm_1$ and $\tz_1$
being nearly pure wino, with $m_{\twpm_1}-m_{\tz_1}\simeq 0.33\,$GeV mass gap. If the mass gap is small enough, then 
charginos can fly a measureable distance before decay. It might be possible to detect 
$e^+e^-\to \twp_1 \twm_1 \gamma \to\gamma+$ soft debris including possible highly ionizing tracks which
terminate into soft pions. 
We presented pMSSM and NMH models with light charginos and sleptons which are in accord with the 
$(g-2)_\mu$ anomaly, $m_h\simeq 125\,$GeV and with a standard neutralino relic abundance 
$\Omega_{\tz_1}^{\mathrm{std}}h^2=0.11$. The ILC-relevant part of the spectrum for benchmark STC 
is very similar to the well-studied SPS1a scenario~\cite{Allanach:2002nj} (or its variant SPS1a').
Finally, we presented the LMH benchmark where the discovered Higgs resonance could turn out to be the
heavier MSSM scalar state $H$ instead of $h$.

In summary, the LHC8 run in 2012 has resulted in the spectacular discover of a SM-like Higgs scalar at 125 GeV.
The Higgs discovery can be regarded as an overall
positive for weak scale supersymmetry in that the mass value falls squarely within the
narrow predicted window predicted by the MSSM. 
However, so far there is no sign of SUSY particles at LHC; instead, 
impressive new limits on gluino and squark masses have been determined. 
Even so, naturalness arguments, possibly  along with the muon $g-2$ anomaly, 
portend a rich assortment of new matter states likely accessible to the ILC, but which remain
difficult to detect at LHC. 
We hope the benchmark models presented here provide a broad picture of the myriad possibilities 
for SUSY physics which may be expected at ILC and also LHC in the post LHC8 era.

\section{Acknowledgments}

We would like to thank Mikael Berggren, Azar Mustafayev, Krzysztof Rolbiecki and Annika Vauth for 
supporting calculations and valuable discussions, and Benno List and Xerxes Tata for 
comments on the manuscript. We thankfully acknowledge the support of the DFG through 
grant SFB 676/2-2010.

\section{Bibliography}


\begin{footnotesize}


\end{footnotesize}


\end{document}